\documentclass[10pt,journal]{IEEEtran}
\ifCLASSINFOpdf
\else
\usepackage{graphicx,amsmath,amssymb,algorithm,algorithmic,tikz,lscape,xspace}
\usepackage[caption=false,font=footnotesize]{subfig}
\graphicspath{{figures/}}
\DeclareGraphicsExtensions{.eps}
\fi
\newcommand{\SP}{\textsc{sp}\xspace}
\renewcommand{\sp}{\SP}
\newcommand{\modsp}{\text{mod}\textsc{sp}\xspace}
\newcommand{\disp}{\textsc{d}\text{i}\textsc{sp}\xspace}
\newcommand{\diomp}{\textsc{d}\text{i}\textsc{omp}\xspace}
\newcommand{\difrogs}{\textsc{d}\text{i}\textsc{frogs}\xspace}
\newcommand{\jsp}{\textsc{jsp}\xspace}
\newcommand{\jomp}{\textsc{jomp}\xspace}
\newcommand{\jfrogs}{\textsc{jfrogs}\xspace}
\newcommand{\omp}{\textsc{omp}\xspace}
\newcommand{\cmp}{\textsc{cmp}\xspace}
\newcommand{\stomp}{\textsc{s}\text{t}\textsc{omp}\xspace}

\newcommand{\modomp}{\text{mod}\textsc{omp}\xspace}
\newcommand{\gp}{\textsc{gp}\xspace}
\newcommand{\digp}{\textsc{d}\text{i}\textsc{gp}\xspace}
\newcommand{\cs}{\textsc{cs}\xspace}

\newcommand{\spursuit}{\textsc{s}\text{-pursuit}\xspace}
\newcommand{\ppursuit}{\textsc{p}\text{-pursuit}\xspace}
\newcommand{\rsupport}{\textsc{r}\text{-support}\xspace}
\newcommand{\isupport}{\textsc{i}\text{-support}\xspace}
\newcommand{\old}{\mathrm{old}}
\newcommand{\cosamp}{\textsc{c}\text{o}\textsc{s}\text{a}\textsc{mp}\xspace}
\newcommand{\resid}{\texttt{resid}}

\newcommand{\maxi}{\texttt{max\_indices}}
\newcommand{\supp}{\texttt{add}_1}
\newcommand{\x}{\mathbf{x}}
\newcommand{\xh}{\hat{\mathbf{x}}}
\newcommand{\y}{\mathbf{y}}
\newcommand{\A}{\mathbf{A}}
\newcommand{\w}{\mathbf{w}}
\renewcommand{\r}{\mathbf{r}}
\newcommand{\T}{\mathcal{T}}
\newcommand{\Tini}{\T_{\text{ini}}}
\newcommand{\Th}{\hat{\mathcal{T}}}
\newcommand{\C}{\mathbf{C}}

\newcommand{\frogs}{\textsc{frogs}\xspace}
\renewcommand{\r}{\mathbf{r}}
\newcommand{\srer}{\textsc{srer}\xspace}
\newcommand{\smnr}{\textsc{smnr}\xspace}
\newcommand{\asce}{\textsc{asce}\xspace}

\newenvironment{packed_enum}{
\begin{enumerate}
  \setlength{\itemsep}{1pt}
  \setlength{\parskip}{0pt}
  \setlength{\parsep}{0pt}
}{\end{enumerate}}

\newtheorem{myfunc}{Function}
\newtheorem{myremark}{Remark}
\newtheorem{proposition}{Proposition}
\newtheorem{definition}{Definition}

\begin{document}
\title{Distributed Greedy Pursuits}
\title{Greedy Pursuits for Distributed Compressed Sensing}
\title{Distributed Greedy Algorithms for \\ Joint Compressed Sensing}
\title{Distributed Greedy Algorithms for \\ Compressed Sensing of Jointly Sparse Signals}
\title{Distributed Greedy Pursuit Algorithms}

% \author{Dennis~Sundman,~\IEEEmembership{Student member,~IEEE,}
%         Saikat~Chatterjee,~\IEEEmembership{Member,~IEEE,}
%         and~Mikael~Skoglund,~\IEEEmembership{Senior Member,~IEEE}
\author{Dennis~Sundman, Saikat~Chatterjee, and Mikael~Skoglund
% <-this % stops a space
%\thanks{M. Shell is with the Department of Electrical and Computer Engineering, Georgia Institute of Technology, Atlanta, GA, 30332 USA e-mail: (see http://www.michaelshell.org/contact.html).}% <-this % stops a space
%\thanks{Manuscript received April 19, 2005; revised January 11, 2007.}}
\thanks{The authors are with Communication Theory Laboratory, School of Electrical Engineering, KTH - Royal Institute of Technology, Sweden. 
Emails: $\{$denniss, sach, skoglund$\}$@kth.se}}

\maketitle

\begin{abstract}
%In distributed compressed sensing, we consider the setup where several sensor nodes are connected through a decentralized (distributed) network; the network can have an arbitrarily connected topology. For this setup, we first introduce a new signal model that helps to exploit correlations between the underlying signals acquired at multiple sensor nodes. Based on this signal model, we then develop new distributed greedy pursuit algorithms for the distributed setup. Incorporating appropriate modifications, we extend the existing orthogonal matching pursuit (\omp) and subspace pursuit (\sp) algorithms for designing two new distributed algorithms. We also develop a new greedy pursuit algorithm which is used for designing a third distributed algorithm. Through simulations, we evaluate the new algorithms and show that the algorithms can provide a performance close to that of a centralized (joint) solution.  
For compressed sensing over arbitrarily connected networks, we consider the problem of estimating underlying sparse signals in a distributed manner. We introduce a new signal model that helps to describe inter-signal correlation among connected nodes. Based on this signal model along with a brief survey of existing greedy algorithms, we develop distributed greedy algorithms with low communication overhead. Incorporating appropriate modifications, we design two new distributed algorithms where the local algorithms are based on appropriately modified existing orthogonal matching pursuit and subspace pursuit. { Further, by combining advantages of these two local algorithms, we design a new greedy algorithm that is well suited for a distributed scenario.} By extensive simulations we demonstrate that the new algorithms in a sparsely connected network provide good performance, close to the performance of a centralized greedy solution.

%We present a brief survey of existing greedy algorithms

%Based on this signal model we develop distributed greedy algorithms with low communication overhead.

%Before we develop the distributed greedy algorithms, we present a brief survey of existing greedy algorithms

%W first designing two new distributed algorithms based on orthogonal matching pursuit and subspace pursuit algorithms for. Then, we develop a new greedy algorithm, 

%which in turn is used to develop a third distributed algorithm. 

\end{abstract}


% \fi
\IEEEpeerreviewmaketitle

\section{Introduction}

\IEEEPARstart{C}{ompressed} sensing (\cs)~\cite{CS:donoho, CRT2006} refers to an under-sampling problem, where few samples of an inherently sparse signal are collected via a linear measurement matrix with the objective of reconstructing the full sparse signal from these few samples. Considering the fact that sparsity is ubiquitous in nature, \cs has many potential applications. In the literature, the task of developing \cs reconstruction algorithms has presumably been considered for a set-up where the samples are acquired by using a single sensor. In the \cs community, we note that there is an increasing effort to consider a multiple-sensor setup.

For a multiple-sensor setup, an interesting case is a distributed setup where several \cs-based sensors are connected through a distributed (decentralized) network. Such a setup is useful in a wide range of applications, for example in distributed sensor perception~\cite{gabs2010} and distributed spectrum estimation~\cite{DistPSD2, DistPSD3, Distributed1}. Considering a camera sensor network, we can envisage a scheme where a set of measurement samples (\cs samples of image signals) from different angles at different positions are acquired. Instead of reconstructing the underlying signals from the corresponding samples independently, one could potentially improve the quality of the reconstructed signals by taking into account all the measurement samples. This is possible by exchanging information over the distributed, but connected network. We refer to this problem as distributed \cs, where the connection between the sensors follows an arbitrary network topology. {Thus, with distributed \cs, we refer to the recovery of a correlated sparse signal where the correlation is in terms of common signal components. If all sensors transmit their measured samples to a common centralized point, the problem can be solved by a centralized algorithm. }For such a setup, we have recently developed joint greedy pursuit reconstruction algorithms in \cite{scs2011}. In the literature, we find that a few more attempts have been made for centralized solutions with various model assumptions \cite{DistCompSens, DistCSjournal1}. Additionally the works based on simultaneous sparse approximation (\textsc{ssa})~\cite{SSA1, SSA2} and multiple measurement vector (\textsc{mmv})~\cite{Distapp1,MMV1} problems, for example simultaneous orthogonal matching pursuit (\textsc{somp}) algorithm~\cite{SOMP}, can be considered to be applied for a centralized (or joint) \cs setup. { The article~\cite{Rakotomamonjy20111505} provides a good overview comparing several centralized algorithms.}

For the distributed \cs setup, we notice some recent attempts to design convex relaxation algorithms~\cite{DistPSD2, DistPSD3, Distributed1,Mota:distributed_basis_pursuit}. { A non-convex algorithmic approach which attempts to minimize a $\ell_q$ minimization problem distributively is presented in~\cite{Qing:decentralized}.} While the convex relaxation algorithms are theoretically elegant and provide good practical performance for low dimensional problems, their use for high dimensional problems are limited due to their high complexity (here, a high dimensional problem refers to the case where the dimensions of underlying signals are high). { Typically the complexity of a convex relaxation algorithm scales with signal dimension $N$ cubically as $\mathcal{O}(N^3)$~\cite{convopt} while for standard \gp algorithms the scaling is $\mathcal{O}(N\log N)$~\cite{Chatterjee12}.} Naturally, designing computationally simple greedy pursuit (\gp), also called greedy search, algorithms is an attractive alternative.
In general, a \gp algorithm uses computationally simple detection and estimation techniques iteratively and hence they are computationally efficient for higher dimensional problems.
While there exists several joint \gp algorithms for the centralized setup, such as \cite{scs2011,SOMP,algo_sim_sparse,MMV1}, there is so far not much attempt for solving the distributed \cs problem based on distributed \gp algorithms. {We first addressed this problem in~\cite{Sundman:a_greedy_pursuit_algorithm} and we found another recent contribution in~\cite{wimalajeewa:cooperative}.}

{In this paper, we develop \gp algorithms for solving the distributed \cs problem where each node reconstructs a signal which is correlated with signals stemming from other sensor nodes.} We refer to the new algorithms as distributed \gp (\digp). {For a distributed \cs setup}, we first introduce a signal model \cite{scs2011} that can describe the correlations between underlying sparse signals. We claim that this new signal model is less restrictive compared to previous signal models~\cite{DistCSjournal1, Distapp1, grsv2008} in the literature.
Based on this signal model, we develop three \digp algorithms. Two of the \digp algorithms are built upon existing \gp algorithms by introducing appropriate modifications. The existing \gp algorithms which we modify are orthogonal matching pursuit (\omp) \cite{OMPfirst} and subspace pursuit (\sp) \cite{SPfirst}. Our motivation for using these two \gp algorithms is that they are good representatives from two main classes of existing \gp algorithms. In the process of using these two \gp algorithms, we realize that there is a scope of developing a new \gp algorithm {by combining advantages from both \omp and \sp which have} high potential for the distributed \cs setup. Hence, we develop a new \gp algorithm {which we call \frogs, followed by its use in the distributed CS setup.}
Through simulations, we evaluate the three new \digp algorithms and show that the algorithms provide increasingly better performance as the network connectivity improves. For a modestly connected network, the simulation results show that the performance is close to the fully connected (centralized) setup and much better than the completely disconnected (independent) setup. In short, the contributions of this paper are:
\begin{itemize}
\item Introduction of a new signal model for solving the distributed \cs problem with correlated data.
\item {Two brief surveys, one on the distributed \cs algorithms and the other on classification of \gp algorithms.}
\item Development of three new distributed greedy pursuit algorithms.
\end{itemize}

%COMMENT: Here add how the paper is arranged.
{ Inspiration for the work in this paper came since the authors were working with improving the performance of \gp algorithm for standard \cs (i.e.~\cite{Chatterjee12}) and from work with the centralized joint sparse signal recovery~\cite{scs2011}.} The remaining parts of the paper are arranged as follows: In the next section, we describe the distributed \cs setup and introduce the new signal model; we also develop a structured approach for describing the quality of connectivity in a distributed network. In Section~\ref{sec:digp}, we introduce the concept of \digp by first studying classifications of different \gp algorithms, and then using this study we develop two \digp algorithms based on existing \omp and \sp. In Section~\ref{sec:dfrogs}, we develop a \gp algorithm  with the aim of providing a \digp algorithm with desirable properties. In Section~\ref{sec:convergence}, we evaluate the convergence of the proposed algorithms. We end the paper with experimental evaluations in Section~\ref{sec:sim_result}.

Notations: Let a matrix be denoted by a upper-case bold-face letter (i.e., $\mathbf{A} \in \mathbb{R}^{M\times N}$) and a vector by a lower-case bold-face letter (i.e., $\mathbf{x} \in \mathbb{R}^{N\times 1}$). $\T$ is the support-set of $\mathbf{x}$, which is defined in the next section. {We also denote $\bar{\T} = \{ 1,2, \dots, N \} \setminus \T$ as the complement to $\T$ where $\setminus$ is the set-minus operator.} $\mathbf{A}_{\T}$ is the sub matrix consisting of the columns in $\mathbf{A}$ corresponding to the elements in the set $\T$. Similarly $\mathbf{x}_{\T}$ is a vector formed by the components of $\mathbf{x}$ that are indexed by $\T$. We let $(.)^{\dagger}$ and $(.)^{T}$ denote pseudo-inverse and transpose of a matrix, respectively. We use $\|.\|$ to denote the $l_2$ norm of a vector.

\section{Distributed Compressed Sensing} \label{sec:distributed_compressed_sensing}
Using a general multiple sensor system setup~\cite{DistCSjournal1}, we first describe the distributed \cs problem and then introduce the new signal model. We have recently proposed this signal model in~\cite{scs2011} and referred to it as the mixed support-set model. In the end of this section we also mention network topology and provide some algorithmic notations.

For the distributed \cs problem, observing the $l$'th sensor, we have the sparse signal $\mathbf{x}_l \in \mathbb{R}^{N}$ measured as
\begin{align}
\mathbf{y}_l = \mathbf{A}_l \mathbf{x}_l + \mathbf{w}_l, ~~~~~~ \forall l \in \{1, 2, ..., L \}, \label{eqn:joint_cs}
\end{align}
where $\mathbf{y}_l \in \mathbb{R}^{M}$ is a measurement vector, $\mathbf{A}_l \in \mathbb{R}^{M\times N}$ is a measurement matrix, and $\mathbf{w}_l \in \mathbb{R}^{M}$ is the measurement error. In this setup $M < N$ and hence the system is under-determined. $\mathbf{A}_l$ and $\mathbf{w}_l$ are independent across $l$. The signal vector $\mathbf{x}_l = [x_l(1) \,\, x_l(2), \ldots ]$ has $K_{l}$ non-zero components with a set of indices $\T_l = \{ i : x_l(i) \neq 0 \}$. $\T_l$ is referred to as the support-set of $\mathbf{x}_l$ with cardinality $|\T_l| =  K_{l}$. 

The distributed \cs reconstruction problem strives to reconstruct $\mathbf{x}_l$ for all $l$ by exploiting some shared structure (correlation) defined by the underlying signal model and by exchanging some information over the given network topology.

\subsection{Mixed support-set model}
\label{sec:mixed_support_signal_model}

Now, we describe the mixed support-set signal model with a shared structure where the signal vector $\mathbf{x}_l$ consists of two parts
\begin{align}
\mathbf{x}_l = \mathbf{z}_l^{(c)} + \mathbf{z}_l^{(p)}, ~~~~~~ \forall l \in \{1, 2, ..., L \}. \label{eqn:ourmodel}
\end{align}
In (\ref{eqn:ourmodel}) both $\mathbf{z}_l^{(c)}$ and $\mathbf{z}_l^{(p)}$ have independent non-zero components. 
The superscripts $(c)$ and $(p)$ represent the notion of `common' part and `private' part, respectively.
For the private part $\mathbf{z}_l^{(p)}$ there are $K^{(p)}_l$ non-zero values. The support-set of $\mathbf{z}_l^{(p)}$ is denoted by $\T_l^{(p)}$. For simplicity we assume that, $\forall l\in\{1,2,\ldots, L \}$, the components of $\T_l^{(p)}$ are drawn uniformly from the set $\{ 1, 2, \ldots, N\}$.
For the common part $\mathbf{z}_l^{(c)}$ there are $K^{(c)}$ non-zero components with the constraint that the associated support-set $\mathcal{T}_l^{(c)}$ is shared as $\T_l^{(c)} = \T^{(c)}, \,\, \forall l\in\{1,2,\ldots, L \}$. While the support-set $\T^{(c)}$ is the same (common) to all signals, it is naturally still unknown to the re-constructor\footnote{For easy practical implementation, we assume that the support-set components are uniformly distributed over $\T^{(c)}$, just as for $\T_l^{(p)}$.}. Here we would like to emphasize that although $\T^{(c)}$ is the same for all sensors, the corresponding non-zero values of $\mathbf{z}_l^{(c)}$ are still individual and {possibly} independent among the nodes. For the $l$'th sensor-node, this gives a support-set $\T_l$ for the signal $\x_l$ as 
\begin{align}
\T_l = \T^{(c)} \cup \T_l^{(p)}, ~~~~~~ \forall l \in \{1, 2, ..., L \}. \label{eqn:support_set}
\end{align}
We define $K_{l,\max} = |\T^{(c)}| + |\T_l^{(p)}| = K^{(c)} + K_l^{(p)}$. Note that the support-sets can intersect, so $K_{l, \max} \geq K_l$. { This model allows for independent signal components among the jointly shared signal data from the sensor nodes. In practice the shared components are likely correlated, which is perfectly supported by the model but any such correlation is not assumed by the reconstruction algorithms developed here. In our recent work \cite{Sundman:diprsp}, we dealt with a model that incorporates such correlation.}

Let us compare the mixed support-set model with signal models already present in the literature. { If we let $\mathbf{z}_l^{(c)} = \mathbf{0}$ the model reduces to the standard, disconnected, \cs problem. On the other hand, if we let $\mathbf{z}_l^{(p)} = \mathbf{0}$, we get
\begin{align}
\mathbf{x}_l = \mathbf{z}^{(c)}_l, ~~~~~~ \forall l \in \{1, 2, ..., L \}, \label{eqn:cotter}
\end{align}
which is the common support-set model~\cite{Distapp1, grsv2008} used in, for example magnetoencephalography, and has no individual signal parts at all. 

We now consider the mixed signal model of~\cite{DistCSjournal1}, where $\mathbf{x}_l$ is composed of common and individual parts
\begin{align}
\mathbf{x}_l = \mathbf{z}^{(c)} + \mathbf{z}_l^{(p)}, ~~~~~~ \forall l \in \{1, 2, ..., L \}. \label{eqn:baranuik}
\end{align}
Here $\mathbf{z}^{(c)}$ represents a common sparse signal part and $\mathbf{z}_l^{(p)}$ represents the individual (private) signal part for the $l$'th sensor. Note that $\mathbf{z}^{(c)}$ is fixed for all the data sets.
Comparing~(\ref{eqn:ourmodel}) and~(\ref{eqn:baranuik}) we can say that the new mixed support-set model provides for additional degrees of freedom since it has no constraint on the common signal value components.
}

A natural question is why we use the mixed support-set model~(\ref{eqn:ourmodel}) for developing \digp algorithms. While we note that the signal model is less stringent in the sense of describing a correlation structure, we also find that the signal model allows us to develop a distributed framework by exchanging a limited information. In this distributed framework, we consider that the estimated support-set at each \cs node as the information to exchange with neighboring nodes.

{
\subsection{Applications}
The mixed support-set model has a high generality and hence can be suitable for a wide range of applications. One example is power spectrum density (PSD) estimation~\cite{sucs2010}. One way to do PSD estimation with \cs for the $l$'th node is to find the sparse edge spectrum based on the autocorrelation coefficients of a measured signal $\y_l$. We worked on such a problem in~\cite{sucs2010}, where the final solution is achieved by solving $\r_{y_l}' = \mathbf{\Phi_{II}} \mathbf{G} \mathbf{z}_{x_l}$; here, $\r_{y_l}'$ are components that can be picked from the correlation matrix $\mathbf{R}_{\y_l} = \mathcal{E}\{ \y_l \y_l^* \}$, while $\mathbf{\Phi_{II}}$ is a transform of the measurement matrix and $\mathbf{G}$ is the inverse of some sparsifying matrix. The distributed CS problem has a direct analogy with the mentioned PSD estimation problem where multiple sensors are used - of course with the presence of measurement noise. 

 Some other examples which can also be cast as a distributed \cs problem include
%Now, we provide few examples of potential applications where the common support-set model can be used: spectrum estimation - where each node experience large overlapping supports in the spectrum~\cite{sucs2010},
multiple sensor image/sound capturing - where each node  observes/listens same object/sound from slightly different angles~\cite{Kirmani:codac,Wu:spherical_microphone}.}
%multiple sensor image capturing - where each node observes the same object from slightly different angles~\cite{Kirmani:codac}, and multiple sensor sound capturing~\cite{Wu:spherical_microphone}. %For all the above scenarios and including the one-sensor scenario, if a slowly varying signal is tracked over time, the proposed mixed support-set model may also apply.
%Since the model only requires the support-set information, this helps us to design distributed \gp algorithms with low communication overhead.

\subsection{Network Topology} \label{sec:network}
In a distributed setup, we assume that the \cs nodes are connected via a network where there is at least one path between any two nodes; otherwise the setup is equivalent to two, or more, independent networks. An example of a simple network can be illustrated by a circular topology where each node (or each sensor) is only connected with another node through a one-way connection (see \figurename~\ref{fig:network_1}). We will refer to this as the worst-case {connected} network of degree 1 and denote it by a connection matrix called $\C_1$ {(observe that this is only a worst-case connected network if we use one-way connections)}. By forming this circular topology of nodes and adding new connections from each node to the others in a systematic way, we can study how the overall performance of a \digp algorithm improves as the network connectivity increases. In \figurename~\ref{fig:network_2}, we show a network where each node is connected to two other nodes (referred to as a degree 2 network) and we denote this network by the connection matrix $\C_2$. In the experimental evaluation (Section~\ref{sec:sim_result}), 
we will study the performance for all intermediate networks, $\C_0, \C_1, ..., \C_{L-1}$ (recall that $L$ is the total number of nodes in the network), where $\C_0$ denotes the use of standard \gp algorithms in a disconnected setup. For the remainder of the paper, if not explicitly stated, $\C_2$ is assumed as the default case for all the \digp algorithms. We will refer to the solution of a fully connected network ($\C_{L-1}$) as the joint solution which is also equivalent to a centralized solution.

The aforementioned network topology approach provides for a systematic analysis of the performance, but to generalize we also work with random networks. The associated notation is $\mathbf{C}_{l, \text{rand}}$ for $l \geq 2$. By this notation we begin with $\mathbf{C}_1$ in~\ref{fig:network_1}. Then, instead of systematically adding connections to the other nodes, we let each node be connected to $l-1$ \textit{random} nodes. This means that there will always be $l$ outgoing connections for each node, but the input can come from any number of incoming nodes less than or equal to $L-1$.

{ To provide a justification related to practical applications, we performed experiments for a 100-node network based on the Watts-Strogatz~\cite{watts1998collective} network model. We also mention that although we generally are interested in a limited and bounded communication overhead for sensor networks, it is outside the scope of this paper to consider potential impact of communication costs corresponding to different networks and algorithms. %This is studied in other works, for example~\cite{Dutta:datamining}
}

\usetikzlibrary{shapes,arrows}
\tikzstyle{line} = [draw, -latex']
\begin{figure}[t]
  \linespread{0.5}
  \centering
  \subfloat[Network of degree 1]{
    \resizebox{0.43\columnwidth}{!}{
      \begin{tikzpicture}[auto]
        \coordinate (a) at (0:15mm);
        \coordinate (b) at (1*36:15mm);
        \coordinate (c) at (2*36:15mm);
        \coordinate (d) at (3*36:15mm);
        \coordinate (e) at (4*36:15mm);
        \coordinate (f) at (5*36:15mm);
        \coordinate (g) at (6*36:15mm);
        \coordinate (h) at (7*36:15mm);
        \coordinate (i) at (8*36:15mm);
        \coordinate (j) at (9*36:15mm);
        \draw [fill] (a) circle (0.5mm);
        \draw [fill] (b) circle (0.5mm);
        \draw [fill] (c) circle (0.5mm);
        \draw [fill] (d) circle (0.5mm);
        \draw [fill] (e) circle (0.5mm);
        \draw [fill] (f) circle (0.5mm);
        \draw [fill] (g) circle (0.5mm);
        \draw [fill] (h) circle (0.5mm);
        \draw [fill] (i) circle (0.5mm);
        \draw [fill] (j) circle (0.5mm);
        \draw [line] (a) -> (b);
        \draw [line] (b) -> (c);
        \draw [line] (c) -> (d);
        \draw [line] (d) -> (e);
        \draw [line] (e) -> (f);
        \draw [line] (f) -> (g);
        \draw [line] (g) -> (h);
        \draw [line] (h) -> (i);
        \draw [line] (i) -> (j);
        \draw [line] (j) -> (a);
      \end{tikzpicture}
      \label{fig:network_1}
    }
  }
  \qquad
  \subfloat[Network of degree 2]{
    \resizebox{0.43\columnwidth}{!}{
      \begin{tikzpicture}[auto]
        \draw [fill] (a) circle (0.5mm);
        \draw [fill] (b) circle (0.5mm);
        \draw [fill] (c) circle (0.5mm);
        \draw [fill] (d) circle (0.5mm);
        \draw [fill] (e) circle (0.5mm);
        \draw [fill] (f) circle (0.5mm);
        \draw [fill] (g) circle (0.5mm);
        \draw [fill] (h) circle (0.5mm);
        \draw [fill] (i) circle (0.5mm);
        \draw [fill] (j) circle (0.5mm);
       \draw [line] (a) -> (b);
        \draw [line] (b) -> (c);
        \draw [line] (c) -> (d);
        \draw [line] (d) -> (e);
        \draw [line] (e) -> (f);
        \draw [line] (f) -> (g);
        \draw [line] (g) -> (h);
        \draw [line] (h) -> (i);
        \draw [line] (i) -> (j);
        \draw [line] (j) -> (a);
        \draw [line,dotted] (a) -> (c);
        \draw [line,dotted] (b) -> (d);
        \draw [line,dotted] (c) -> (e);
        \draw [line,dotted] (d) -> (f);
        \draw [line,dotted] (e) -> (g);
        \draw [line,dotted] (f) -> (h);
        \draw [line,dotted] (g) -> (i);
        \draw [line,dotted] (h) -> (j);
        \draw [line,dotted] (i) -> (a);
        \draw [line,dotted] (j) -> (b);
      \end{tikzpicture}
      \label{fig:network_2}
    }
  }
  \caption{Network topologies of degree 1 ($\C_1$) and degree 2 ($\C_2$).}
  \label{fig:network}
\end{figure}
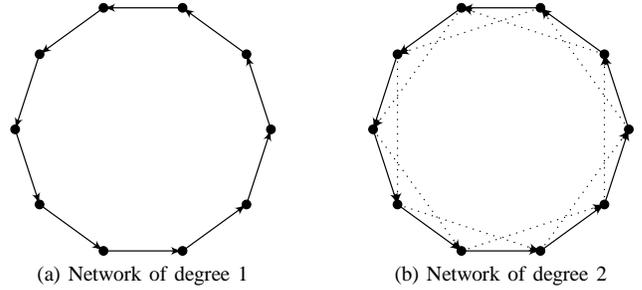

\subsection{Algorithmic notation}
For clarity in the algorithmic notation, we define three functions as follows:
\begin{align} \label{eqn:resid}
\resid(\mathbf{y}, \mathbf{B}) \triangleq \mathbf{y} - \mathbf{B} \mathbf{B}^{\dagger} \mathbf{y},
\end{align}
where $\mathbf{y}$ is a vector and $\mathbf{B}$ is a full column-rank matrix;
\begin{align} \label{eqn:max}
{\maxi}&(\mathbf{x}, k) \triangleq \{ \textit{the set of indices corresponding} \nonumber \\
& \hspace{-10mm}\textit{to the $k$ largest amplitude components of } \mathbf{x} \},
\end{align}
and
\begin{align} \label{eqn:add}
\supp(\mathbf{s}, \T) & \triangleq \{ \forall j \in \T, \textit{ perform } s_{j} = s_{j}+1 \},
\end{align}
where $\mathbf{s}=[s_{1} \,\, s_{2} \,\, \ldots s_{N} ]$ and $s_{j} \geq 0$.

For the $l$'th \cs node, $\mathcal{L}_{l}^{\text{out}}$ denotes the set of indices corresponding to the outgoing connected nodes and $\mathcal{L}_{l}^{\text{in}}$ denotes the set of indices corresponding to the incoming connected nodes (we always consider that the $l$'th \cs node is connected with itself and hence $\mathcal{L}_{l}^{\text{in}}$ and $\mathcal{L}_{l}^{\text{out}}$ have at least an element that corresponds to the $l$'th node itself).

{
\subsection{A Literature Survey of Distributed \cs Algorithms}
We now present a short survey of distributed \cs algorithms already present in the literature. First, we endeavor to distinguish between a distributed/centralized solver and the distributed/centralized \cs problem. A solution algorithm can be either distributed or centralized independent of whether the underlying signals to be estimated are correlated or not. For example, the standard, one-sensor, \cs problem can be solved by a distributed algorithm~\cite{Pattersson:distributed,Mota:distributed_basis_pursuit}. In this paper we concern ourselves with the case where the distributed \cs problem (with correlated signal measurements) is solved by a distributed algorithm.

%These concepts are independent of each other, but for a distributed \cs problem it can be argued that it is of interest to provide a distributed solver. In this paper, we limit our interest to the distributed solvers. %In this section, we distinguish between convex and greedy solvers.

%The convex solvers are all based on the common signal model, see \eqref{eqn:baranuik}. This means that each sensor in a sensor network measures, possible with different sensing-matrices and different measurement noise, the same (or parts of the same) signal.a

\subsubsection{Convex Solvers}
For the sensor node $l$, the convex solvers are of the form
\begin{align}
  \underset{\xh_l}{\min} \| \xh_l \|_1 ~ ~ \text{such that} ~ ~ \| \y_l - \A_l\xh_l \|_2 \leq \epsilon_l, \label{eqn:bpdn}
\end{align}
which is often referred to as the basis pursuit denoising problem. They can also take another form, called the Lasso problem
\begin{align}
  \underset{\xh_l}{\min} \| \xh_l \|_1 + \lambda \| \y_l - \A_l \xh_l \|_2. \label{eqn:lasso}
\end{align}

The distributed convex algorithms solves the problem where $\x_l = \x$ which means the objective is for each node to reach the same solution.
The distributed basis pursuit~\cite{Mota:distributed_basis_pursuit} solves \eqref{eqn:bpdn} for two different signal models by re-writing the problem on a form so that a distributed optimization method called alternating direction method of multipliers (ADMM)~\cite{Boyd:2011:DOS:2185815.2185816} can be applied. The D-Lasso algorithm~\cite{DistPSD3} solves \eqref{eqn:lasso} considering a specific application scenario, mainly PSD estimation in time, frequency and space. D-Lasso also has to find a consensus for $\lambda$ which is done in parallel with solving the \cs problem. The BPDN problem can be solved with the simplex algorithm~\cite{dantzig1965linear}, which can also solved in a distributed manner, shown in~\cite{Dutta:datamining,Yarmish:2001:DIS:933505,Stunkel:1989:HIS:63047.63104}. 

\subsubsection{Greedy Solvers}
The \gp algorithms attempt to solve the \cs problem which gives a strictly $K_{\max}$-sparse solution. Recently some attempts have been made to solve the distributed \cs problem with distributed \gp algorithms.

The distributed and collaborative \omp (\textsc{dc-omp}) algorithm in~\cite{wimalajeewa:cooperative} is an extension to \omp. The \textsc{dc-omp} algorithm is similar to the \diomp algorithm presented in this paper, but instead of waiting for the inner algorithm to finish, it exchanges and decides on which components to be added after each iteration of the inner algorithm. This algorithm is based on the assumption that each node wants to reconstruct the same signal and can not work with the mixed support-set model.

%In Appendix~\ref{app:modOMP}, we provide a modified version of \omp in Algorithm~\ref{alg:OMP}, which reduces to regular \omp when $\Tini = \emptyset$. If we let $\Tini = \emptyset$ in Algorithm~\ref{alg:OMP}, we can explain the \textsc{dc-omp} by introducing a communication and decision phase between step~\ref{OMP:selmax} and \ref{OMP:union}. In the communication phase, the index selected in step~\ref{OMP:selmax} is exchanged among the neighbors and in the decision phase, one or several indices are chosen among the received ones.

Solving the iterative hard thresholding (IHT)~\cite{blumensath2009iterative} problem in a distributed manner is done in~\cite{Pattersson:distributed} by the algorithm D-IHT. Here, two computations are done, a local and a global execution. The global execution has to be performed by a central node, which requires a much stricter network infrastructure than a regular distributed algorithm.

We developed a distributed predictive \sp algorithm in~\cite{Sundman:diprsp} that is based on the common support-set model with correlated coefficients. This algorithm uses the predictive \sp algorithm previously developed in~\cite{zachariah:dip}. Iteratively employing the neighbors' signal estimates, which are exchanged in a distributed network, signal and covariance priors are formed which are used in the predictive \sp algorithm.

If we compare the work proposed in this paper with the above works, we notice that the algorithms presented in this article are based on the less restrictive signal model \eqref{eqn:ourmodel}. Furthermore, the algorithms presented here are fully distributed with no need for a centralized node.% We provide the algorithms without any analytical bounds but with extensive experimental evaluation.
}
\section{Distributed Greedy Pursuits} \label{sec:digp}

In this section, we develop two different \digp algorithms based on two existing \gp algorithms. Furthermore, in Section~\ref{sec:dfrogs} we develop a new \gp algorithm on which we construct the third \digp algorithm. The three \digp algorithms that are developed in this paper are referred to as follows:
\begin{enumerate}
 \item Distributed \omp (\diomp): Where we use existing orthogonal matching pursuit (\omp) \cite{OMPfirst} as the \gp algorithm after appropriate modifications.
 \item Distributed \sp (\disp): Where we use existing subspace pursuit (\sp) \cite{SPfirst} as the \gp algorithm after appropriate modifications.
 \item Distributed \frogs (\difrogs): Where we use new forward-reverse orthogonal greedy search (\frogs) as the \gp algorithm. The \frogs algorithm is described in Section~\ref{sec:frogs}.
\end{enumerate}
For these three \digp algorithms, we find that it is possible to develop distributed algorithmic structures in two different ways. The \diomp follows the first distributed algorithmic structure, and the \disp and \difrogs follows the second distributed algorithmic structure. These two distributed algorithmic structures are developed in Section \ref{sec:domp} and \ref{sec:dsp} where they are developed as two examples of the \digp algorithms, \diomp and \disp, respectively.
However, for developing the \digp algorithms, we first need to know preliminaries about underlying \gp algorithms. This helps to bring appropriate modifications to the \gp algorithms or construct new \gp algorithms, so that they are better suited for the development of \digp algorithms. A brief survey of \gp algorithms is presented in the following section.

\subsection{A Brief Survey of \gp Algorithms}  

In general, for CS reconstruction, existing \gp algorithms are used with an implicit assumption of a single-sensor setup. Using the measurement vector collected from the sensor, the main principle of the \gp algorithms is to estimate the underlying support-set of a sparse vector followed by valuating the associated signal values. The support-set is the set of indices corresponding to the non-zero elements of a sparse vector. To estimate the support-set and the associated signal values, the \gp algorithms use linear algebraic tools, for example the matched filter detection and least-squares estimation. 
A crucial point worth mentioning is that the success of the \gp algorithms mainly depends on their efficiency in estimating the support-set. Once a support-set is formed, the associated signal values can be obtained by a simple least-squares estimation.
 
In the literature, we note two main algorithmic approaches for the \gp algorithms: 1) the categorization of \textit{serial} or \textit{parallel}, and 2) the construction mechanism in \textit{reversible} or \textit{irreversible} manner. First let us consider the algorithmic approach of serial or parallel support-set construction strategy. If serial construction is performed then elements of the support-set are chosen one-by-one; in contrast, for parallel construction, several elements of the support-set are chosen simultaneously. Next we consider the algorithmic approach of reversible and irreversible construction. If irreversible construction is performed then an element already added to the support-set, remains there indefinitely; in contrast, for reversible construction, an element of the support-set (chosen in the past) can be removed later (if the element is found to be unreliable). Therefore, considering serial or parallel construction, a \gp algorithm can be categorized either as a serial pursuit (\spursuit) or a 
parallel pursuit (\ppursuit) algorithm. On the other hand, considering reversible or irreversible, a \gp algorithm can either use a reversible support-set (\rsupport) or an irreversible support-set (\isupport) construction mechanism.  

We categorize several \gp algorithms in Table~\ref{table:categories} where we consider existing \omp \cite{OMPfirst}, \sp \cite{SPfirst}, \cosamp \cite{netro2009}, look ahead orthogonal least-squares (\textsc{laols}) \cite{Chatterjee12}, stagewise omp (\stomp) \cite{donoho2006}, backtracking \omp (\textsc{baomp}) \cite{backtrackingomp}, projection-based \omp(\textsc{pomp}) \cite{Chatterjee12}, look ahead parallel pursuit (\textsc{lapp}) \cite{swectwsuch}, regularized \omp (\textsc{romp}) ~\cite{Needell:signal_recovery_incomplete}, cyclic matching pursuit (\textsc{cmp}) \cite{CMP_2007, CMP_2011}, and the new forward-reverse orthogonal greedy search (\frogs) algorithm. For developing \digp algorithms, we use the \omp and \sp among the existing \gp algorithms because they are generic and easy to implement. We develop the \frogs algorithm since it seems promising for a distributed setup. The development of the \frogs algorithm and its use in constructing a \digp algorithm are reported in Section~\ref{sec:dfrogs}.
Now, for developing \digp algorithms based on the signal model (mixed support-set model~(\ref{eqn:ourmodel})) and the algorithmic architectures of \gp algorithms, we find the principle strategies discussed in the following section.

%For developing \digp algorithms, we use the \omp and \sp among the existing \gp algorithms because they are generic and easy to implement. Furthermore, to the best of our knowledge, we are unable to find an existing \gp algorithm in the current literature which can be categorized as \spursuit algorithm with the characteristics of \rsupport construction mechanism. 
%Such an algorithm is particularly suited for a distributed framework. 

%Hence, we develop the new \frogs algorithm to fill the gap in Table~\ref{table:categories}. The development of the \frogs algorithm and its use in constructing a \digp algorithm are reported in Section~\ref{sec:dfromp}.

%Now, for developing \digp algorithms based on the signal model (mixed support-set model~(\ref{eqn:ourmodel})) and the algorithmic architectures of \gp algorithms, we find the principle strategies discussed in the following section. 

\begin{table}[!t]\renewcommand{\arraystretch}{1.3}
\caption{Classification of \gp algorithms}
\label{table:categories}
\centering
\begin{tabular}{|c||c|c|} \hline
 & $\ppursuit$ & $\spursuit$ \\ \hline \hline
$\rsupport$ & \sp, \cosamp, \textsc{lapp}, \textsc{baomp} & \textsc{cmp}, $\frogs$ \\
\hline
\isupport &  \stomp, \textsc{romp} & \omp, \textsc{laols}, \textsc{pomp}\\
\hline
\end{tabular}
\end{table}

\subsection{Principle Strategies for \digp Algorithms}
The new iterative \digp algorithms are developed based on two principle strategies which are invoked in each iteration of the algorithms. The two principle strategies are described below:
\begin{enumerate}
 \item Each \cs node transmits its own full support-set estimate to the neighboring connected nodes. It also receives a set of full support-set estimates from the neighboring connected nodes.
 \item Using the set of all received support-set estimates and by invoking a voting mechanism, each \cs node finds an estimate of the common support-set, either serially or parallelly. Then, using the common support-set estimate as the initial knowledge, each \cs node finds a new estimate of the full support-set and then again exchange the full support-set information.    
\end{enumerate}
Using the two principles, the \digp algorithms continue to execute until convergence.
Now, considering \omp and \sp as the underlying \gp algorithms, we describe the voting mechanism and develop two \digp algorithms in the next sections. Later, in Section~\ref{sec:dfrogs}, we develop the third \digp algorithm based on \frogs.

\subsection{Voting: Find the Common Support-set}
Based on a number of full support-set estimates, a significant task in the distributed \cs problem based on the common support-set model is to find an estimate of the common support-set. It seems clear that if a certain index is present in all full support-set estimates from the incoming nodes, this index is a strong candidate for being part of the common support-set. Thus, a consensus vote (i.e., the intersection) among the support-sets would be a prominent approach. However, in practice it turns out that due to errors in support-set estimates, a consensus is not always possible. Instead, as often is the case when a consensus cannot be reached, majority voting is a prominent approach. Thus we develop a method which chooses the common support-set to be the set of indices which overlaps with most full support-set estimates from the incoming nodes (c.f., majority voting). This approach is shown in Algorithm~\ref{alg:voting}.
\begin{algorithm}[ht!]
\caption{: Voting based choice of indices \newline
\textit{Executed in $l^*$-th node, where $\mathcal{L}_{l^*}^{\text{in}}$ is the set of neighboring incoming nodes (Note that $l^* \in \mathcal{L}_{l^*}^{\text{in}}$)}} \label{alg:voting} 
{Input: $\mathcal{L}_{l^*}^{\text{in}}$, $\{ \hat{\T}_{l} \}$ where $l \in \mathcal{L}_{l^*}^{\text{in}}$, and the desired cardinality $K^{\text(c)}$ of common support set}\vfill
\begin{algorithmic}[1]
\STATE $\mathbf{s}_{l^*} \leftarrow \mathbf{0}_{N\times 1}$ 
\FOR { each $l \in \mathcal{L}_{l^*}^{\text{in}}$}
\STATE $\mathbf{s}_{l^*} \leftarrow \supp(\mathbf{s}_{l^*}, \Th_l)$
\ENDFOR
\STATE $\hat{\T}_{l^*}^{(c)} \leftarrow \maxi(\mathbf{s}_{l^*}, K^{\text(c)})$ \hfill (Note: $| \hat{\T}_{l^*}^{(c)} |= K^{\text(c)} $)
\end{algorithmic}
\mbox{Output: $\hat{\T}_{l^*}^{(c)}$}
\end{algorithm}
In Algorithm~\ref{alg:voting}, we supply the inputs: $\mathcal{L}_{l^*}^{\text{in}}$, $\{ \hat{\T}_{l} \}$ where $l \in \mathcal{L}_{l^*}^{\text{in}}$, $K^{\text(c)}$. Here, $\mathcal{L}_{l^*}^{\text{in}}$ denotes the neighboring incoming nodes, $\{ \hat{\T}_{l} \}$ is the estimated support-sets in all the connected nodes, and $K^{\text(c)}$ is the desired cardinality of the common support-set. The output of Algorithm~\ref{alg:voting} is the common support-set estimate $\hat{\T}_{l^*}^{(c)}$; here we use the subscript $l^*$ to denote the case for $l^*$-th node. Then, using the $\supp(., .)$ and $\maxi(., .)$ functions, the voting algorithm finds the common support-set estimate. In this case, we rely on the fact that the elements of the common support-set have the highest scores in terms of their occurrences. Hence, the method can be viewed as a democratic voting strategy. Using Algorithm~\ref{alg:voting}, we define the following function.
\begin{myfunc}
\label{func:Voting_based_choice_of_indices}
(Voting based choice of indices) For the $l^*$-th node, let the set of neighboring incoming nodes $\mathcal{L}_{l^*}^{\text{in}}$, the estimated support-sets $\{ \hat{\T}_{l} \}$ in the incoming nodes such that $l \in \mathcal{L}_{l^*}^{\text{in}}$, and the desired cardinality $q$ of the common support-set be given. Then, the estimated common support-set $\hat{\T}_{l^*}^{(c)}$ is the output of the following algorithmic function 
\begin{align*}
\hat{\T}_{l^*}^{(c)} \leftarrow \texttt{vote} (\mathcal{L}_{l^*}^{\text{in}}, \{ \hat{\T}_{l} \}, q  ). 
\end{align*}
where the above function executes Algorithm~\ref{alg:voting}.
\end{myfunc}

\subsection{Distributed \omp} \label{sec:domp}

For developing the distributed \omp (\diomp) algorithm, we first modify the existing \omp algorithm and then use the modified \omp algorithm as a building block. The modified \omp is referred to as \modomp where the modification is required so that it can use an initial support-set estimate in its task of estimating the full support-set. The \modomp is presented in \ref{app:modOMP}. 

Now, we consider the $l^*$-th node and develop the \diomp algorithm based on \modomp and the voting function. \diomp is shown in Algorithm~\ref{alg:disupport} and it is executed locally and distributively in each node of the connected network. Let us consider \diomp for the $l^*$-th node.
\begin{algorithm}[ht!]
\caption{: Distributed $\omp$ (\diomp) \newline
\textit{Executed in $l^*$-th node, where $\mathcal{L}_{l^*}^{\text{in}}$ and $\mathcal{L}_{l^*}^{\text{out}}$ is the set of incoming and outgoing nodes, respectively ($l^* \in \mathcal{L}_{l^*}^{\text{in}}$, $l^* \in \mathcal{L}_{l^*}^{\text{out}}$).}} \label{alg:disupport}
\mbox{Input: $\mathbf{A}_{l^*}$, $\mathbf{y}_{l^*}$, $K^{(p)}_{l^*}$, $K^{(c)}$}\newline
\mbox{Initialization: }
\begin{algorithmic}[1]
\STATE $K_{l^*,\max} = K^{(p)}_{l^*} + K^{(c)}$
\STATE $(\hat{\T}_{l^*}, \hat{\mathbf{x}}_{l^*}, \eta_{l^*}) \leftarrow \modomp(\mathbf{A}_{l^*}, K_{l^*,\max}, \mathbf{y}_{l^*}, \emptyset) $
\STATE $\Th_{l} \leftarrow \emptyset, ~ \forall ~ l \in \mathcal{L}_{l^*}^{\text{in}} \setminus l^*$ \hfill (i.e. except $l^*$)
\STATE $k \leftarrow 0$ \hfill (iteration counter)
\end{algorithmic}
\mbox{Iteration:}
\begin{algorithmic}[1]
\REPEAT
\STATE $k \leftarrow k + 1$
\STATE \{ \textbf{Transmit:} $\hat{\T}_{l^*}$ to all nodes $l \in \mathcal{L}_{l^*}^{\text{out}}$ \} \label{disupport:send}
\STATE \{ \textbf{Receive:}  $\hat{\T}_l$ from all nodes $l \in \mathcal{L}_{l^*}^{\text{in}}$ \}\label{disupport:recv}
\STATE $\hat{\T}_{l^*}^{(c)} \leftarrow \texttt{vote} (\mathcal{L}_{l^*}^{\text{in}}, \{ \hat{\T}_{l} \}, k )$ \label{disupport:s_end} \hfill (Note: $| \hat{\T}_{l^*}^{(c)} |=k$)
\STATE $(\hat{\T}_{l^*}, \hat{\mathbf{x}}_{l^*}, \eta_{l^*}) \leftarrow \modomp(\mathbf{A}_{l^*}, K_{l^*,\max}, \mathbf{y}_{l^*}, \hat{\T}_{l^*}^{(c)}) $ \label{disupport:gp}
\UNTIL{$k = K^{(c)}$}
\end{algorithmic}
\mbox{Output: $\hat{\mathbf{x}}_{l^*}$, $\hat{\T}_{l^*}$}
\end{algorithm}
The \emph{input} to \diomp (Algorithm~\ref{alg:disupport}) is the $l^*$-th node's sensing matrix $\A_{l^*}$, the measurement vector $\y_{l^*}$, and the cardinality of the private and common support-sets $K_{l*}^{(p)}$ and $K^{(c)}$.
For the \emph{initialization} phase, before any communication has taken place, $\modomp(\cdot)$ is executed to achieve a first estimate of the $l^*$-th node's support-set. At this phase, all incoming neighboring support-set estimates (where the incoming neighboring nodes are identified by $\mathcal{L}_{l^*}^{\text{in}} \setminus l^*$) are initialized as empty sets $\emptyset$ and an iteration parameter $k$ is initialized to zero.
The \emph{iteration} phase of \diomp is characterized by three main functionalities: 1)~In steps~\ref{disupport:send} and \ref{disupport:recv}, the communication phase takes place, where the support-set estimates are exchanged among the nodes. Note that the $l^*$-th node transmits its estimated support-set $\hat{\T}_{l^*}$ to the all outgoing nodes indexed by $\mathcal{L}_{l^*}^{\text{out}}$ and also receives the support-set estimates $\{ \hat{\T}_l \}$ from the incoming nodes.
2)~In step~\ref{disupport:s_end}, by using all the full support-sets estimates $\{ \hat{\T}_l \}$ the voting strategy is invoked to achieve an estimate of the common support-set. Note that the intermediate common support-set is estimated in each iteration and its cardinality is increased one-by-one through iterations (serially). 
3)~Using the estimated common support-set, $\modomp(\cdot)$ is executed in step~\ref{disupport:gp} to achieve a new full support-set estimate together with a signal estimate for the $l^*$-th node. These three functionalities are iteratively executed until the common support-set cardinality becomes $K^{(c)}$. Therefore, the \diomp algorithm iterates exactly $K^{(c)}$ times.

In \diomp, it is worth mentioning the importance of the serial construction mechanism strategy for the common support-set estimation. 
For compressible sparse signal vectors, where the sorted amplitudes of the signal vectors quickly decays (for example, if the non-zero components of a sparse signal vector is drawn from an i.i.d. Gaussian source), it is known that the serial construction is more efficient \cite{Chatterjee12}. Hence, to estimate the common support-set reliably, we use the serial construction. However, the serial construction requires more computation in practice and we endeavor for developing a parallel construction mechanism with less complexity.

\subsection{Distributed \sp} \label{sec:dsp}

We now develop the second \digp algorithm using a parallel support-set construction mechanism. The new \digp approach is based on the existing \sp algorithm \cite{SPfirst} and hence referred to as distributed \sp (\disp). Like \diomp, we first modify the \sp algorithm (we refer to the modified \sp as \modsp which is explained in \ref{app:modSP}) and then use it for developing \disp.

\disp is shown in Algorithm~\ref{alg:drsupport}, where we use the voting function of Algorithm~\ref{alg:voting} and \modsp of Algorithm~\ref{alg:SP}. The principle strategy in Algorithm~\ref{alg:drsupport} is the same as that of \diomp; the strategy is to improve the common support-set estimation by exchanging full support-set estimates over iterations. 
In each iteration of \disp, the common support-set estimate $\Th_{l^*}^{(c)}$ is passed to the \modsp algorithm which in turn finds the full support-set estimate $\hat{\T}_{l^*}$. Using the voting mechanism, we here find the $\Th_{l^*}^{(c)}$ with full cardinality ($|\Th_{l^*}^{(c)}|=K^{(c)}$) in each iteration. This kind of parallel common support-set construction may allow for a faster convergence than the serial common support-set construction used in \diomp. Here, we mention that the parallel common support-set construction for \disp is realizable with high reliability because we use \modsp, which has an reversible construction mechanism (i.e., \modsp may remove bad elements of support-set in a later iteration). We now take a closer look on \disp in Algorithm~\ref{alg:drsupport}.

\begin{algorithm}[ht!]
\caption{: Distributed \sp (\disp) \newline \textit{Executed in $l^*$-th node, where $\mathcal{L}_{l^*}^{\text{in}}$ and $\mathcal{L}_{l^*}^{\text{out}}$ is the set of incoming and outgoing nodes, respectively ($l^* \in \mathcal{L}_{l^*}^{\text{in}}$, $l^* \in \mathcal{L}_{l^*}^{\text{out}}$).}}\label{alg:drsupport}
\mbox{Input: $\mathbf{A}_{l^*}$, $\mathbf{y}_{l^*}$, $K^{(p)}_{l^*}$, $K^{(c)}$}\newline
\mbox{Initialization: }
\begin{algorithmic}[1]
\STATE $K_{l^*,\max} = K^{(p)}_{l^*} + K^{(c)}$
\STATE $(\hat{\T}_{l^*}, \hat{\mathbf{x}}_{l^*}, \eta_{l^*}) \leftarrow \modsp(\mathbf{A}_{l^*}, K_{l^*,\max}, \mathbf{y}_{l^*}, \emptyset) $
\STATE $\eta_{l^*}^{\old} \leftarrow \eta_{l^*}$
\STATE $\Th_{l} \leftarrow \emptyset, ~ \forall ~ l \in \mathcal{L}_{l^*}^{\text{in}} \setminus l^*$ \hfill (i.e. except $l^*$)
\end{algorithmic}
\mbox{Iteration:}
\begin{algorithmic}[1]
\REPEAT
\IF {$ \eta_{l^*} > \eta_{l^*}^{\old} $} \label{dsp:resif}
\STATE $(\hat{\T}_{l^*}, \hat{\mathbf{x}}_{l^*}, \eta_{l^*}) \leftarrow (\hat{\T}_{l^*}^{\old}, \hat{\mathbf{x}}_{l^*}^{\old}, \eta_{l^*}^{\old})$
\ENDIF \label{dsp:resendif}
\STATE $(\hat{\T}_{l^*}^{\old}, \hat{\mathbf{x}}_{l^*}^{\old}, \eta_{l^*}^{\old}) \leftarrow (\hat{\T}_{l^*}, \hat{\mathbf{x}}_{l^*}, \eta_{l^*})$
\STATE $\hat{\T}_{l}^{\old} \leftarrow \hat{\T}_{l}, ~ \forall ~ l \in \mathcal{L}_{l^*}^{\text{in}} \setminus l^*$
\STATE \{ \textbf{Transmit:} $\hat{\T}_{l^*}$ to all nodes $l \in \mathcal{L}_{l^*}^{\text{out}}$ \} \label{dsp:send}
\STATE \{ \textbf{Receive:} $\hat{\T}_l$ from all nodes $l \in \mathcal{L}_{l^*}^{\text{in}}$ \}\label{dsp:recv}
\STATE $\hat{\T}_{l^*}^{(c)} \leftarrow \texttt{vote} (\mathcal{L}_{l^*}^{\text{in}}, \{ \hat{\T}_{l} \}, K^{(c)} )$ \label{drsupport:s_end} \hfill (Note: $| \hat{\T}_{l^*}^{(c)} |=K^{(c)}$) \label{dsp:s_end}
\STATE $(\hat{\T}_{l^*}, \hat{\mathbf{x}}_{l^*}, \eta_{l^*}) \leftarrow \modsp(\mathbf{A}_{l^*}, K_{l^*,\max}, \mathbf{y}_{l^*}, \hat{\T}_{l^*}^{(c)}) $ \label{dsp:SPI}
\UNTIL{$((\eta_{l^*} \geq \eta_{l^*}^{\old})$ \AND $(\hat{\T}_l = \hat{\T}_l^{\old}, ~ \forall ~ l \in \mathcal{L}_{l^*}^{\text{in}}))$}\label{dsp:convergence}
\end{algorithmic}
\mbox{Output: $\hat{\mathbf{x}}_{l^*}^{\text{old}}$, $\hat{\T}_{l^*}^{\text{old}}$}
\end{algorithm}

\emph{Input} to Algorithm~\ref{alg:drsupport} is the $l^*$-th node's sensing matrix $\mathbf{A}_{l^*}$, the measurement vector $\y_{l^*}$, and the cardinality of the private and common support sets, i.e., $K^{(p)}_{l^*}$ and $K^{(c)}$. In the \emph{initialization} phase of the algorithm, before any communication has taken place, $\modsp(\cdot)$ is executed to achieve a first estimate of the $l^*$-th node's support-set. The residual norm $\eta_{l^*}$ is stored to use as the performance measure and the support-set estimates of the neighboring nodes are initialized as the empty set $\emptyset$. In the \emph{iteration} phase of \disp, there are four main functionalities:
1) Steps~\ref{dsp:resif}~to~\ref{dsp:resendif} prevent the result from deviating away from a better solution, which empirically was observed to happen if the intermediate estimated support-set in step~\ref{dsp:SPI} was worse than the estimated support-set in the previous iteration (denoted by the use of `old'). 2) Steps~\ref{dsp:send}~to~\ref{dsp:recv} constitute the communication phase, where the locally estimated support-sets are exchanged among the connected nodes. 3) Using the voting function (of Algorithm~\ref{alg:voting}) in step~\ref{dsp:s_end}, an estimate of the common support-set with full cardinality $K^{(c)}$ is achieved. 4) Using the estimated common support-set, $\modsp(\cdot)$ is executed locally to estimate a new full support-set, to again be communicated over the network. These four functionalities are iteratively performed until convergence is achieved. For convergence, we have a stopping criterion based on two conditions to be fulfilled together: a) the residual norm in the $l^*$-th 
node does not decrease, and b) no new support-set estimates from connected nodes arrive. The second condition is used due to the fact that if the $l^*$-th node receives new improved support-set estimates from its neighbors then its own support-set estimate may improve in a later iteration.

\subsection{Further Scope of Improvement}  \label{subsec:distributed_sp}

Different strategies are used for developing \diomp and \disp algorithms. For \diomp, we build the common support-set estimate $\Th^{(c)}$ serially. The \diomp is built on the use of \modomp which is categorized as an \spursuit algorithm with the characteristics of \isupport construction mechanism (see Table~\ref{table:categories}). In \modomp, the use of serial approach (\spursuit) allows for a high reliability to detect the correct element in the current iteration, but (still for \modomp) the irreversible support-set construction mechanism (\isupport) also has a disadvantage. The disadvantage is that if an incorrect element is found to be reliable and added to the support-set in a previous iteration then the element remains in the support-set forever. 
In contrast, for \disp, we use the parallel approach where the common support-set estimate (with full cardinality) is refined iteratively. The \disp is built on the use of \modsp and the \modsp is categorized as a \ppursuit algorithm with characteristics of the \rsupport construction mechanism. The \rsupport construction mechanism has the capability to remove a wrong element in a future iteration even though the element was found to be reliable and added to the support-set in a past iteration. This support-set construction mechanism is reversible in nature (thus the notation \rsupport). 

Considering the advantages of the serial approach (\spursuit) and the reversible construction mechanism (\rsupport), we develop a new \gp algorithm in the next section that has both the characteristics. By considering Table~\ref{table:categories}, we notice that \cmp~\cite{CMP_2007,CMP_2011} is a prominent candidate already fulfilling these two characteristics. However, \cmp requires an iteration parameters to be provided. This parameter tells the algorithm how many times it should search its current support-set estimate for replacing indices. In a distributed scenario where the voting algorithm is used, it may happen that the entire support-set estimate is completely wrong. For \cmp to correct for this kind of scenario, we need the iteration parameter to be in the order of $K^{(c)}$ which poses computational burden. While developing \digp algorithms this computational burden turned out prohibitive. Thus, we develop the new \frogs algorithm to be serial, reversible and better suited for our needs, and use it 
as a building block to develop a third \digp scheme.

\section{Distributed \frogs} \label{sec:dfrogs}

In this section, we first develop the new \gp algorithm called forward-reverse \omp (\frogs) which is an \spursuit algorithm with \rsupport characteristic. Based on \frogs, we then develop a new \digp algorithm called distributed \frogs (\difrogs).

\subsection{\frogs} \label{sec:frogs}

The development of \frogs is based on \omp. For \omp, a careful study reveals that the use of highest-amplitude based element-selection strategy leads to a natural selection scheme in which the elements are chosen in an ordered manner. Ideally \omp serially detects the elements according to their decreasing strength of amplitudes. The success of this ordered selection strategy depends on the level of system uncertainty. For a highly under-sampled system, the highest amplitude based selection strategy may fail to detect a correct element and erroneously include a wrong element in the support-set. To improve this strategy further, a reliability testing procedure after the selection may be helpful for eliminating the errors. 

For developing \frogs, we refer the serial add strategy of including a potential element in the support-set as \textit{forward add}. This forward add strategy is directly used in standard \omp and we present it as a separate algorithm in Algorithm~\ref{alg:fa}.
\begin{algorithm}[ht!]
\caption{: Forward-add}\label{alg:fa}
\mbox{Input: $\mathbf{A}$, $\r_k$, $\T_k$}
\begin{algorithmic}[1]
  \STATE  $\tau_{\max} \leftarrow \maxi(\A^T \r_k, 1)$ \label{fa:forward_start}
  \STATE  $\T_{k+1} \leftarrow \T_{k} \cup \tau_{\max}$
  \STATE  $\r_{k+1} \leftarrow \resid(\y, \A_{\T_{k+1}})$ \label{fa:forward_end}
\end{algorithmic}
\mbox{Output: $\r_{k+1}$, $\T_{k+1}$}
\end{algorithm}
In Algorithm~\ref{alg:fa}, we supply the inputs: $\mathbf{A}$, $\r_k$, $\T_k$. Here, $\mathbf{A}$ is the sensing matrix, $\r_k$ is the residual vector for iteration $k$ and $\T_k$ is the support-set estimate for iteration $k$. Then, analogous to \omp, $\maxi(.,.)$ and $\resid(.,.)$ are used and the algorithm outputs the residual $\r_{k+1}$ and support-set $\T_{k+1}$ for iteration $k+1$. Now using Algorithm~\ref{alg:fa}, we define the following function.
\begin{myfunc}
(Forward-add) For the $k$'th iteration, the sensing matrix $\A$, the current residual $\r_k$ and the current support-set $\T_{k}$ are given. Then, for the $(k+1)$'th iteration, the new support-set with cardinality $(|\T_k|+1)$ and its corresponding residual are the outputs of the following algorithmic function 
\begin{align*}
(\r_{k+1}, \T_{k+1}) \leftarrow \texttt{forward\_add}(\mathbf{A}, \r_k, \T_k),
\end{align*}
which exactly executes Algorithm~\ref{alg:fa}.
\end{myfunc}
After the forward-add strategy is performed, it is natural to include a reliability testing strategy. For this, we develop a new scheme where the $k$ most prominent support-set elements are chosen from $(k+1)$ elements. This new selection algorithm is presented in Algorithm~\ref{alg:rf}.
\begin{algorithm}[ht!]
\caption{: Reverse-fetch}\label{alg:rf}
\mbox{Input: $\mathbf{A}$, $\T_{k+1}$, $k$} \hfill \mbox{(Note: $|\T_{k+1}|=k+1$)}
\vspace{-11pt} %TODO: REMOVE THIS if possible
\begin{algorithmic}[1]
  \STATE $\tilde{\x}$ ~ such that ~ $\tilde{\x}_{\T_{k+1}} = \mathbf{A}_{\T_{k+1}}^{\dagger} \mathbf{y}$ and $\tilde{\x}_{\bar{\T}_{k+1}} = \mathbf{0}$ \label{frogs:reverse_project}
  \STATE    $\T' \leftarrow \maxi(\tilde{\x}, k)$   \hfill (Note: $|\T'|=k$) \label{frogs:reverse_maxi}
  \STATE    $\r' \leftarrow \resid(\y, \A_{\T'})$ \label{frogs:reverse_resid}
\end{algorithmic}
\mbox{Output: $\r'$, $\T'$}
\end{algorithm}
In Algorithm~\ref{alg:rf}, we supply the inputs $\mathbf{A}$, $\T_{k+1}$ and $k$. By using least squares estimation, we find an estimate of the intermediate $k+1$ non-zero elements of the sparse signal $\tilde{\x}$. Based on this signal estimate, the temporary support-set $\T'$ of cardinality $k$ and the corresponding temporary residual $\r'$ are found. Using Algorithm~\ref{alg:rf} we define the following function.
\begin{myfunc}(Reverse-fetch) Let the sensing matrix $\A$ and a support-set $\T_{k+1}$ of cardinality $k+1$ be given. Then the temporary support-set $\T'$ with cardinality $k$ and its corresponding residual are the outputs of the following algorithmic function
\begin{align*}
(\r', \T') \leftarrow \texttt{reverse\_fetch}(\A, \T_{k+1}, k),
\end{align*}
which exactly executes Algorithm~\ref{alg:rf}.
\end{myfunc}

\begin{algorithm}[ht!]
\caption{: Forward-Reverse orthogonal greedy search ($\frogs$)}\label{alg:frogs}
\mbox{Input: $\mathbf{A}$, $K_{\max}$, $\mathbf{y}$, $\T_{\text{ini}}$} \newline
\mbox{Initialization: }
\begin{algorithmic}[1]
\STATE $\mathbf{R}=\left[ \mathbf{r}_{1} \,\, \mathbf{r}_{2} \,\, \ldots, \mathbf{r}_{K_{\max}} \right]$  \hfill (For storing residuals)
\STATE $(\T_0, \x_0) \leftarrow \modomp(\y, \A, K_{\max}, \T_{\text{ini}})$ \label{frogs:modomp}
\FOR{$l = 1:K_{\max}$} \label{frogs:sort_start}
\STATE $\T' \leftarrow \maxi(\x_0, l)$  \hfill (Note: $| \T' |=l$)
\STATE $\r_l \leftarrow \resid(\y, \A_{\T'})$
\ENDFOR \label{frogs:sort_end}
\STATE $k \leftarrow K_{\max}$, $\T_k \leftarrow \T'$
\end{algorithmic}
\mbox{Iteration:}
\begin{algorithmic}[1]
  \REPEAT
\STATE $(\r_{k+1}, \T_{k+1}) \leftarrow \texttt{forward\_add}(\mathbf{A}, \r_k, \T_k)$ \label{frogs:forward}
  \REPEAT \label{frogs:reverse_repeat}
\STATE $(\r', \T') \leftarrow \texttt{reverse\_fetch}(\A, \T_{k+1}, k)$ \label{frogs:reverse}
  \IF{$(\| \r' \|  < \| \r_k \| )$} \label{frogs:reverse_check} %% Go back step 
  \STATE      $\T_{k} \leftarrow \T'$, $\r_{k} \leftarrow \r'$ \label{frogs:reverse_reverse_start}
  \STATE      $k \leftarrow k - 1$ \label{frogs:reverse_reverse_end}
  \ELSE
  \STATE      break
  \ENDIF \label{frogs:reverse_check_end}
  \UNTIL{ $(k = 0)$ } \label{frogs:reverse_end}
  \STATE  $k \leftarrow k + 1$ \label{frogs:reverse_deiterate}
  \UNTIL{$k = K_{\max}+1$}
\end{algorithmic}
\mbox{Output: }
\begin{algorithmic}[1]
\STATE $\Th = \T_{k-1}$
\STATE $\xh$ ~ such that ~ $\xh_{\Th} = \mathbf{A}_{\Th}^{\dagger} \mathbf{y}$ and $\xh_{\bar{\Th}} = \mathbf{0}$
\STATE $\eta = \|\r_{k-1}\|$
\end{algorithmic}
\hrule
\mbox{Functional form:} $(\hat{\T}, \hat{\mathbf{x}}, \eta) \leftarrow \frogs(\mathbf{A}, K, \mathbf{y},\T_{\text{ini}}) $
\end{algorithm}

\begin{figure*}
{
  \linespread{0.5}
  \centering
  \subfloat[Inner loop. Comparison for \newline varying $\alpha$'s.]{
    \resizebox{0.49\columnwidth}{!}{
      \includegraphics[width=\columnwidth]{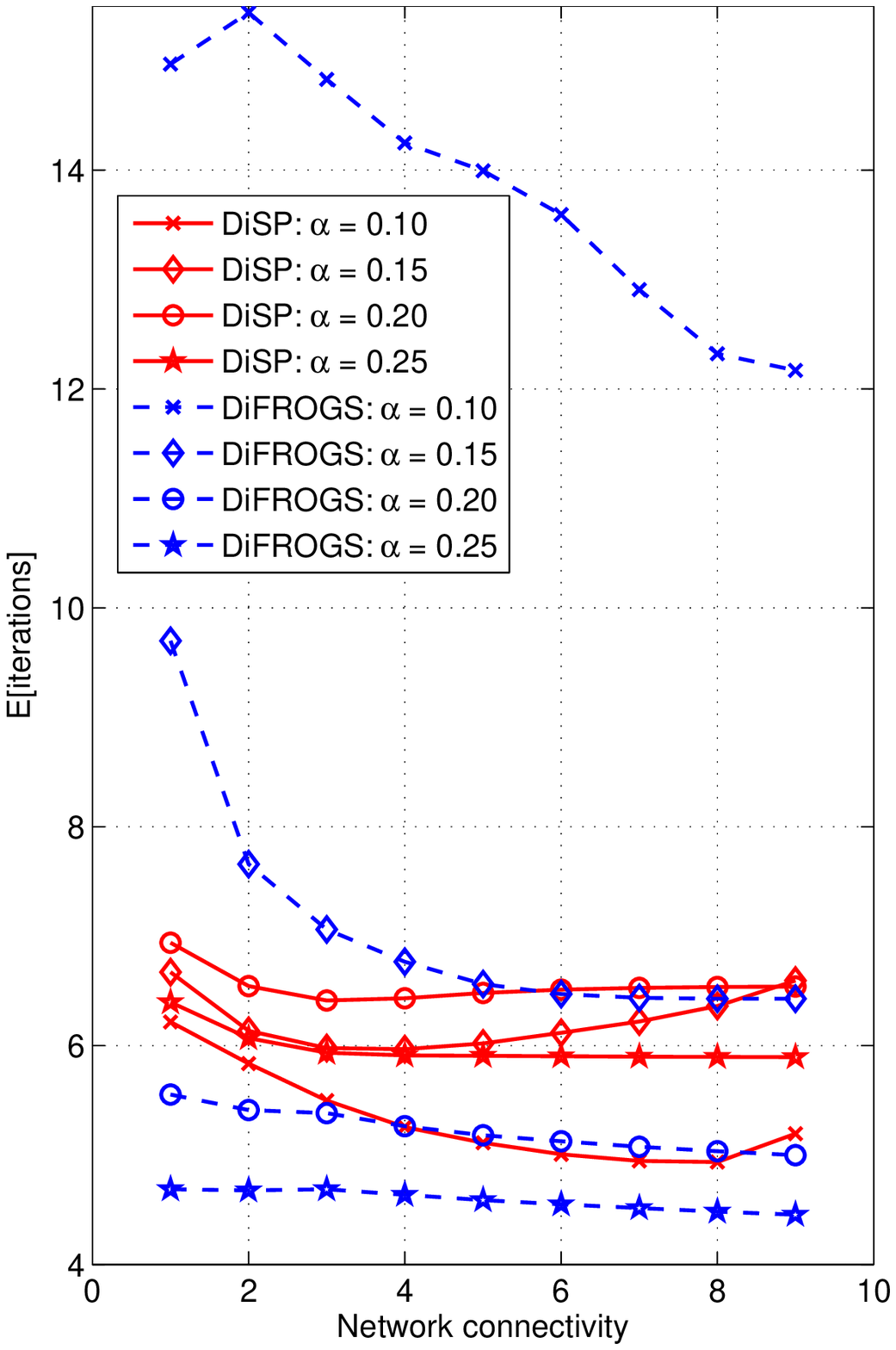}
      \label{fig:inner_loop_all}
    }
  }
  \subfloat[Inner loop. Here $\alpha = 0.15$. \newline Showing one standard deviation.]{
    \resizebox{0.49\columnwidth}{!}{
      \includegraphics[width=\columnwidth]{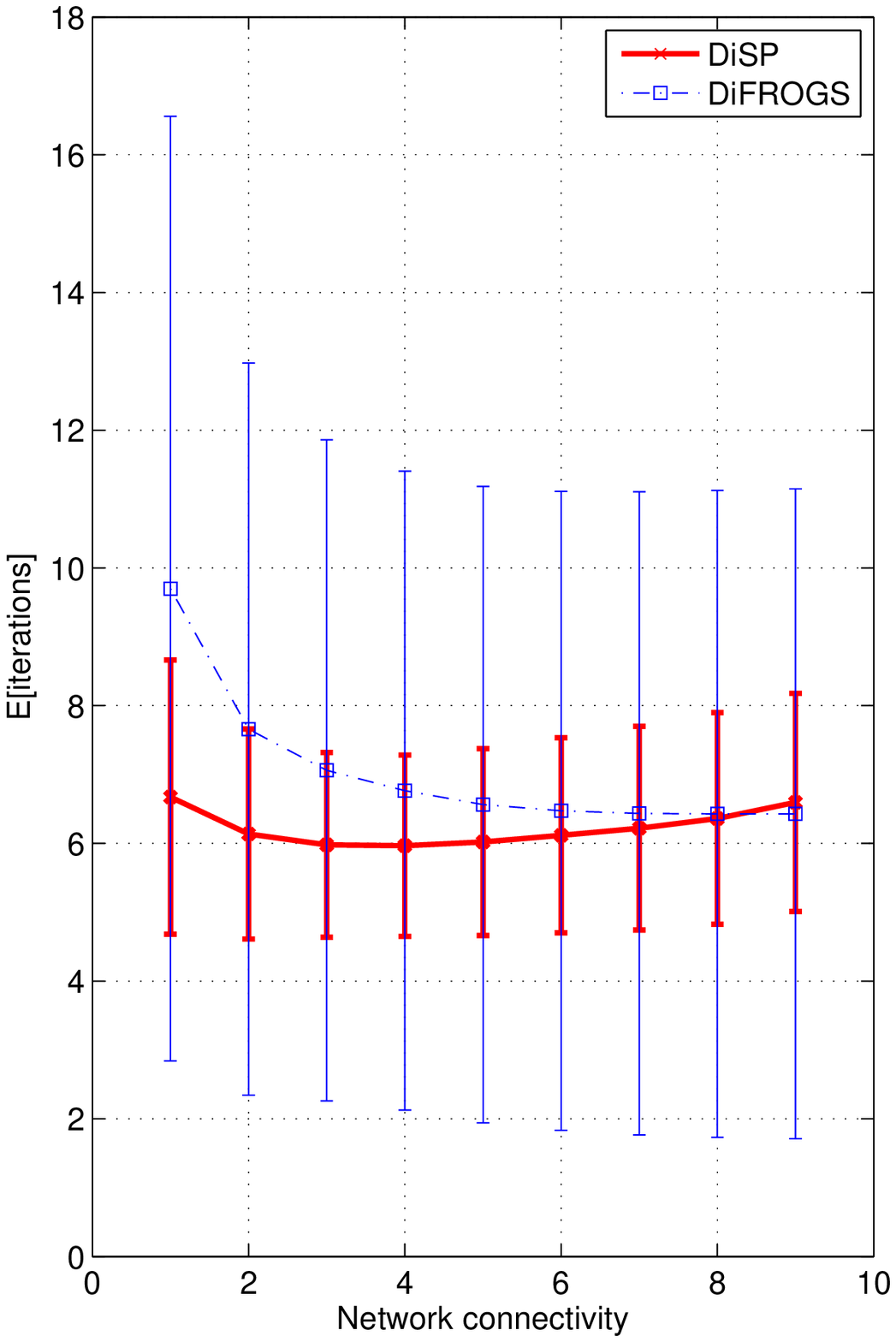}
      \label{fig:inner_loop_two}
    }
  }
  \subfloat[Outer loop. Comparison for \newline varying $\alpha$'s.]{
    \resizebox{0.49\columnwidth}{!}{
      \includegraphics[width=\columnwidth]{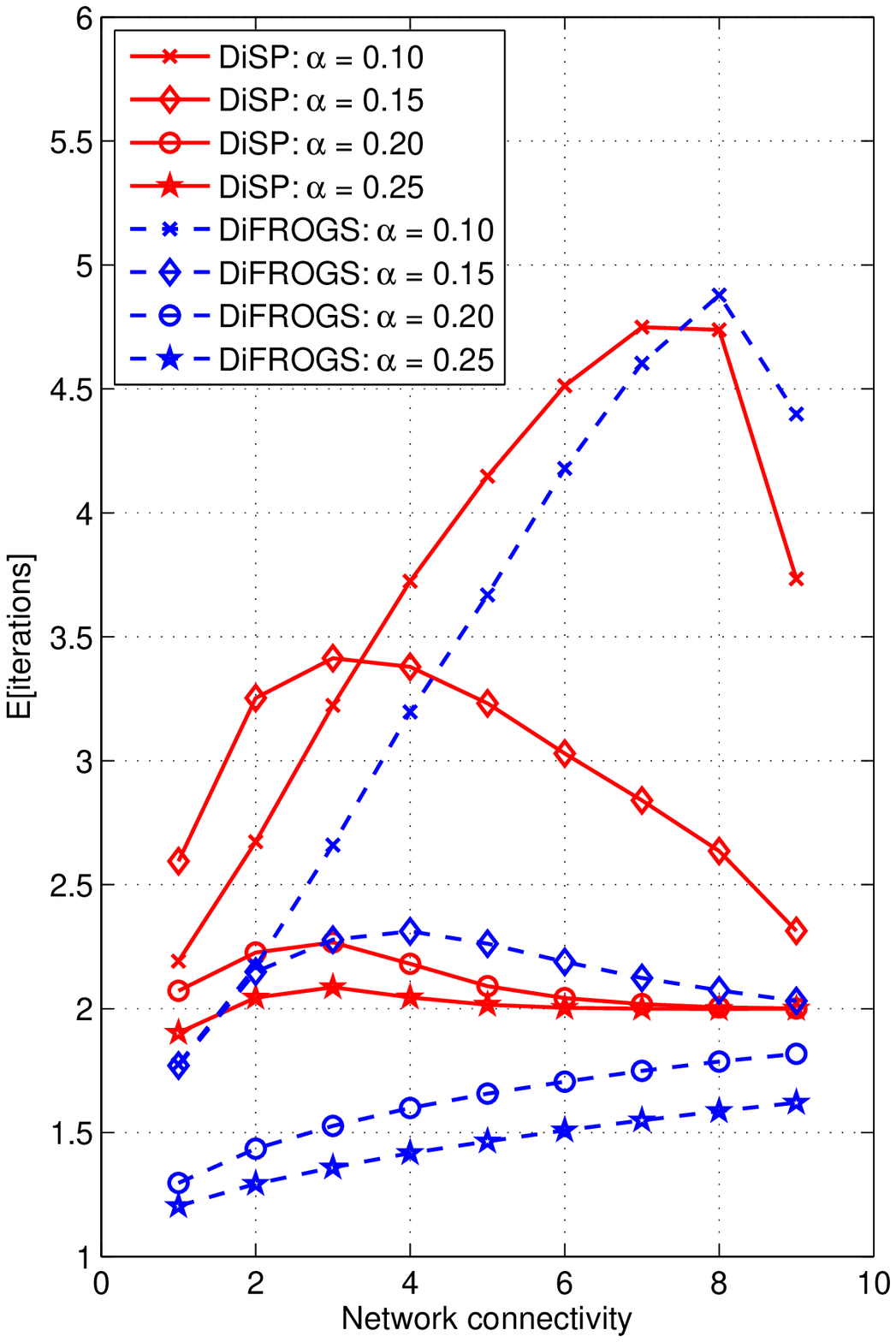}
      \label{fig:outer_loop_all}
    }
  }
  \subfloat[Outer loop. Here $\alpha = 0.15$. \newline Showing one standard deviation.]{
    \resizebox{0.49\columnwidth}{!}{
      \includegraphics[width=\columnwidth]{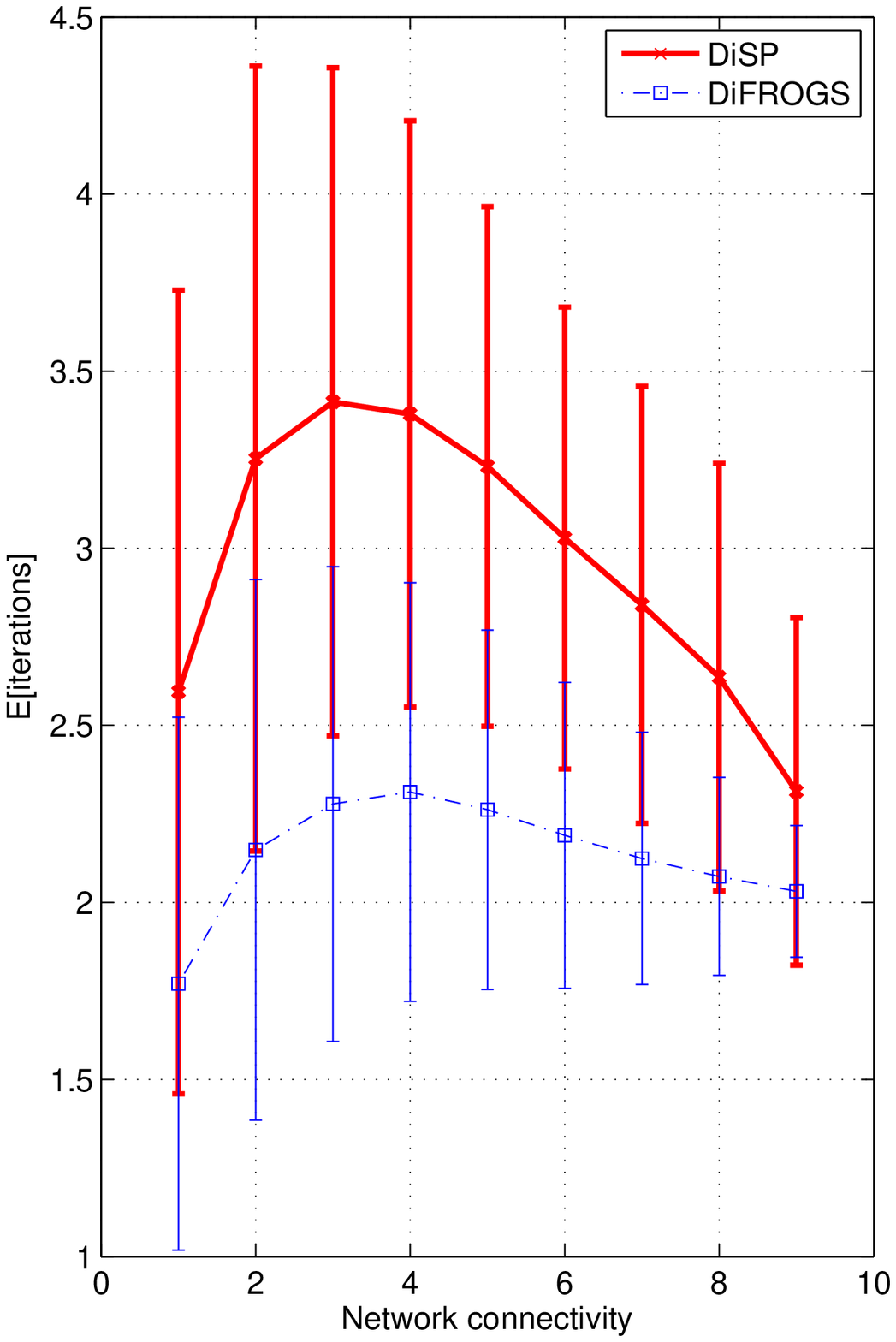}
      \label{fig:outer_loop_two}
    }
  }

  \caption{The average number of iterations for the inner and outer loops. Comparison between $\disp$ and $\difrogs$ for several different $\alpha$'s. Also showing one standard deviation in the case where $\alpha = 0.15$.}
  \label{fig:iterations}
}
\end{figure*}

Based on the $\texttt{forward\_add}()$ and $\texttt{reverse\_fetch}()$ functions, we now develop the \frogs in Algorithm~\ref{alg:frogs}.
Similarly to the $\modsp$ and $\modomp$ algorithms, the inputs to \frogs are $\A$, $K_{\max}$, $\y$ and $\Tini$.
In the \textit{initialization} phase \frogs calls the $\modomp$ (see \ref{app:modOMP}) procedure (step~\ref{frogs:modomp}) to form a full support-set estimate $\T_0$. If there are errors in $\Tini$, those errors will remain in $\T_0$ (and in $\x_0$). Then, in steps~\ref{frogs:sort_start} to \ref{frogs:sort_end}, an ordering procedure is performed which helps to arrange the corresponding residual vectors appropriately. This ordering is necessary for the reliability testing.
Notice that the iteration phase starts with $k = K_{\max}$. For clarity, we denote $\r_{k}$, $\r_{k+1}$ and $\r'$ as the current, intermediate and temporary residuals, respectively.  
In the $k$'th \textit{iteration}, two main tasks are performed.
First, when the algorithm performs $\texttt{forward\_add}()$, the output is an intermediate support-set $\T_{k+1}$ with cardinality larger than the current support-set by one. Second, for reliability testing, the $\texttt{reverse\_fetch}()$ function is invoked to find the $k$ elements from the intermediate support-set of cardinality $(k+1)$. These $k$ elements form the temporary support-set $\T'$. Then, considering the residual norm as the model fit measure, a comparison between residual norms is performed. For the comparison, if the temporary residual norm $\| \r' \|$ is smaller than the current residual norm $\| \r_{k} \|$, then the temporary support-set $\T'$ acts as the current new support-set $\T_k$. Similarly if $\| \r' \|$ is smaller than $\| \r_{k} \|$,  $\r'$ replaces $\r_k$. Now, the algorithm decreases the iteration counter by one and continues the reverse operation of refining the support-set. Note that the reverse operation is a serial operation, similar to the forward-add operation. In the case 
when $\| \r' \|$ is not smaller than $\| \r_{k} \|$, the reverse operation is not performed; we assume that the current support-set is reliable and $\texttt{forward\_add}()$ is performed for the inclusion of a new element (serially). As both the operations - the forward operation of increasing the support-set and reverse operation of correcting the support-set - are performed in a serial manner, we conclude that the new \frogs algorithm can be categorized as an \spursuit algorithm with \rsupport construction mechanism.

{
\subsection{Non-rigorous Discussion on the Behavior of \frogs}
%$\texttt{forward\_add}()$ and $\texttt{reverse\_fetch}()$ methodology in
In the serial \modomp algorithm, if an initial support-set contains errors, the errors will not be corrected since it is an irreversible algorithm. Since the \diomp algorithm shows good performance, the initial motivation for \frogs is to create a serial algorithm that can correct such errors reversibly.

We believe that there is also another reason for the success of FROGS. In the FROGS, the $\texttt{forward\_add}()$, which also is a support-set index selection method similar to \omp, chooses the one largest-in-amplitude $\tau_{\max}$ component from $\A^T\r_k$. However, $\r_k$ contains two error terms, see for example (6) in \cite{zachariah:dip}:
\begin{align}
  \r_k = \sum_{i\in\bar{\T}_{k}} \mathbf{a}_i x_i + \sum_{i\in\T_k} \mathbf{a}_i \xi_{k,i} + \w,
\end{align}
where $\xi_{k,i} = x_i - \hat{x}_{k,i}$; here $\xi_{k,i}$ refers to the result of the $k$'th iteration and the $i$'th component. As the support-set estimate improves, the second error term will decrease. Because of this second error term, a support-set index chosen early in the iterations may be erroneously chosen. In the $\texttt{reverse\_fetch}()$, the least squares estimator minimizes
%If this is the case, then in the $\texttt{reverse\_fetch}()$ of a later iteration this index will have a smaller impact on $\xh$ than the currently found term
%The behavior of the $\texttt{forward\_add}()$ will be to add the largest signal-components first. In the $\texttt{reverse\_fetch}()$, the least squares estimator minimizes
\begin{align}
  \underset{\xh_{\T_{k+1}}}{\arg \min} ~ ~ \| \y - \A_{\T_{k+1}} \xh_{\T_{k+1}} \|_2^2. \label{eqn:ls}
\end{align}
The least squares approach estimates the signal $\xh$ based on the full support-set and will therefore be more reliable than using a matched filter. By using the least squares to evaluate the support-set, the algorithm has a chance to detect (and remove) the previously found erroneously chosen index since this index will have a smaller contribution in $\xh$ than an index found in a later iteration.

%that the signal components have been found in a decreasing order, the forward-reverse strategy will improve the performance of \frogs.
%In this equation, the minimization is done over all signal components within the current support-set. In comparison with the matched filter, this means that we this re-evaluation step includes the new component and the error term seen for least squares is smaller.
}

\subsection{Distributed \frogs} \label{subsec:dfrogs}

The distributed \frogs (\difrogs) is designed based on the new \frogs algorithm.
Since \frogs is an \rsupport algorithm, we can develop \difrogs by using the same approach as \disp. In fact, it turns out that by just replacing $\modsp(\cdot)$ with $\frogs(\cdot)$ in Algorithm~\ref{alg:drsupport}, we can develop the \difrogs algorithm.

\begin{myremark}
Following the development of \diomp and \disp based on modified \omp and \sp, respectively, and then developing \difrogs based on new \frogs, we can safely claim that many existing \gp algorithms can be modified and new \gp algorithms can be developed for building new \digp algorithms. For example, we could easily modify \stomp or \cosamp for the purpose of developing new \digp algorithms.\end{myremark}

{
\section{Convergence}\label{sec:convergence}
For the distributed algorithms developed in this paper we consider two iteration parameters, the outer loop iteration parameter (i.e., the parameter for \diomp, \disp and \difrogs) and the inner loop iteration parameter (i.e., the parameter for \modomp, \modsp and \frogs). Here the outer loop parameter refers to how many times a local algorithm runs; and for one algorithm run, the inner loop parameter refers to how many iterations the local algorithm iterates. We present the number of iterations (instead of i.e., time, or floating points per second) as complexity measure since the number of iterations is independent on specific implementation. Based on previous analysis for \omp and \sp we can derive analytical results for \modomp and \modsp, respectively. For the inner and outer loops of \difrogs and \disp we also provide a numerical evaluation. Since both inner and outer loop of \diomp turns out to be fixed, there is no reason to numerically evaluate their convergence.

In the numerical results, we show the average number of iterations with a confidence interval of one standard deviation. We have used $N = 500$, $K^{(c)} = 10$, $K^{(p)} = 10$. To measure the level of under-sampling, we define the fraction of measurements
\begin{align}
\alpha = \frac{M}{N}. \label{eqn:alpha}
\end{align}

\subsection{Convergence of \diomp}
In \diomp, the number of inner and outer iterations is constant by construction. Therefore, we can exactly determine the total number of inner iterations. 
\subsubsection{Inner Iterations: Iterations for \modomp}
Since \modomp is a modification of \omp in such a way that the algorithm, from the initial support $\Tini$, continues to build a support-set estimate until the estimate is of size $K_{\max}$, we find that the number of iterations for the inner loop of \diomp (i.e., \modomp) is exactly $K_{\max} - |\Tini|$. This follows directly from the construction of the stopping criterion in \modomp. 

\subsubsection{Outer Iterations: Iterations for \diomp} 
By construction it is clear that the outer loop of \diomp will run exactly $K^{(p)}$ iterations.

%Thus, the total number of iterations for \diomp is $K^{(p)} \left( K_{\max} - |\Tini| \right)$. In comparison, the disconnected \omp algorithm always (in the noisy case) iterates $K_{\max}$ times.

\subsection{Convergence of \disp}
For the inner loop, since \modsp is based on \sp, we can use tools developed in~\cite{giryes:sptrans} and \cite{giel2010tr} to analyze the convergence. The outer loop is instead evaluated by numerical experiments.

\subsubsection{Inner Iterations: Iterations for \modsp}
From previous work~\cite{SPfirst,giryes:sptrans} we know that \sp fulfill certain performance criteria. In particular we mention
\begin{proposition}[Theorem 2.1 in~\cite{giryes:sptrans}, Corollary 3.2 in~\cite{giel2010tr}] \label{prop:sp} For a $K$-sparse vector $\x$, under the condition $\delta_{3K} \leq 0.139$, the solution of \sp at the $k$th iteration satisfies
  \begin{align}
    \| \x_{\bar{\T}_k} \|_2 \leq 2^{-k} \| \x \|_2 + 16.44 \| \A^T_{\T_{\w}} \w \|_2.
  \end{align}
  In addition, after at most $k_{\sp}^* = \left\lceil \log_2\left( \frac{\| \x \|_2}{\| \A^T_{\T_{\w}} \w \|_2} \right)\right\rceil$ iterations, the solution $\xh_{\sp}$ leads to an accuracy
  \begin{align}
    \| \x - \xh_{\sp} \|_2 \leq 21.41 \| \A^T_{\T_{\w}} \w \|_2.
  \end{align}
\end{proposition}
Here, $\T_{\w}$ is the support-set of size $K$ corresponding to the columns in $\A$ that are most strongly correlated with the noise, $\T_{\w}~=~\underset{\T}{\arg \max} \| \A^T_{\T} \w \|_2$. For \modsp, we can form a similar proposition with the difference that the initial support-set enters into the expression.
\begin{proposition} \label{prop:modsp}
  For a $K$-sparse vector $\x$, under the condition $\delta_{3K} \leq 0.139$, the solution of \modsp at the $k$th iteration satisfies
  \begin{align}
    \| \x_{\bar{\T}_k} \|_2 \leq 2^{-k} \| \x_{\bar{\T}_{\text{ini}}} \|_2 + 16.44 \| \A^T_{\T_{\w}} \w \|_2.
  \end{align}
  In addition, after at most $k_{\modsp}^* = \left\lceil \log_2\left( \frac{\| \x_{\bar{\T}_{\text{ini}}} \|_2}{\| \A^T_{\T_{\w}} \w \|_2} \right)\right\rceil$ iterations, the solution $\xh_{\modsp}$ leads to an accuracy
  \begin{align}
    \| \x - \xh_{\modsp} \|_2 \leq 21.41 \| \A^T_{\T_{\w}} \w \|_2.
  \end{align}
  \begin{IEEEproof}
    See Appendix~\ref{app:modSP}
  \end{IEEEproof}
\end{proposition}

From Proposition~\ref{prop:sp} and \ref{prop:modsp} we draw the conclusion that \modsp will reach the same performance as \sp but in fewer iterations (i.e., $k^*_{\modsp} \leq k^*_{\sp}$), since $\| \x_{\bar{\T}_{\text{ini}}} \|_2 \leq \|  \x \|_2$.

These propositions provide worst-case bounds which are loose~\cite{giryes:sptrans,giel2010tr}. In practice, it may be more useful to study the numerical evaluation of the average number of iterations $\bar{k}$ in \figurename~\ref{fig:inner_loop_all} and \ref{fig:inner_loop_two}. In \figurename~\ref{fig:inner_loop_all}, we see that \modsp provide an average number of iterations between 5-7. Also by \figurename~\ref{fig:inner_loop_two}, we notice from the standard deviation that the number of iterations are less varying than for \difrogs.

\subsubsection{Outer Iterations: Iterations for \disp}
We intuitively expect the number of outer iterations for \disp to vary with the network connectivity. By studying \figurename~\ref{fig:outer_loop_all},  we see that this is also the case. However, it is unexpected to see that (for all $\alpha$'s), the average number of iterations do not change consistently with network connectivity. Instead, we notice that the maximum number of iterations occur for $\alpha = 0.15, 0.20$ at $\C_3$, for $\alpha = 0.10$ at $\C_8$ and for $\alpha = 0.25$ at $\C_9$. %[TODO: provide a reason for this?]

\subsection{Convergence of \difrogs}
The new \difrogs algorithm is developed using pure engineering intuitions, resulting in several `if-else' statements which are hard to analyze theoretically. Instead of theoretical analysis we study the numerical evaluation in \figurename~\ref{fig:iterations}.

\subsubsection{Inner Iterations: Iterations for \frogs}
In \figurename~\ref{fig:inner_loop_all}, we see that the inner iterations for \difrogs varies depending on $\alpha$. In particular, for $\alpha = 0.10$, we see that the number of inner iterations is significantly higher than for all other cases. This behavior is a result of the uncertainty caused by having so few measurements at hand that the reverse-fetch procedure often activates to correct for errors. Except for this extreme case, the number of iterations seems to be comparable to \disp. Furthermore, as the network connectivity increases, the number of iterations decreases which makes sense since the common support-set estimate will be better. In \figurename~\ref{fig:inner_loop_two}, we notice that the average number of iterations for \difrogs are more fluctuating than for \disp.

\subsubsection{Outer Iterations: Iterations for \difrogs}
In \figurename~\ref{fig:outer_loop_all}, we see that the outer iterations for \difrogs varies similarly to \disp. For the bigger $\alpha = 0.20$ and $\alpha = 0.25$, the number of iterations are consistently fewer for \difrogs compared to \disp. By studying \figurename~\ref{fig:outer_loop_two}, we see that as the network connectivity increases, the uncertainty in the average number of iterations becomes smaller.

}

\section{Performance Results} \label{sec:sim_result}
\begin{figure*}
  \linespread{0.5}
  \centering
  \subfloat[$\diomp$]{
    \resizebox{0.66\columnwidth}{!}{
      \includegraphics[width=\columnwidth]{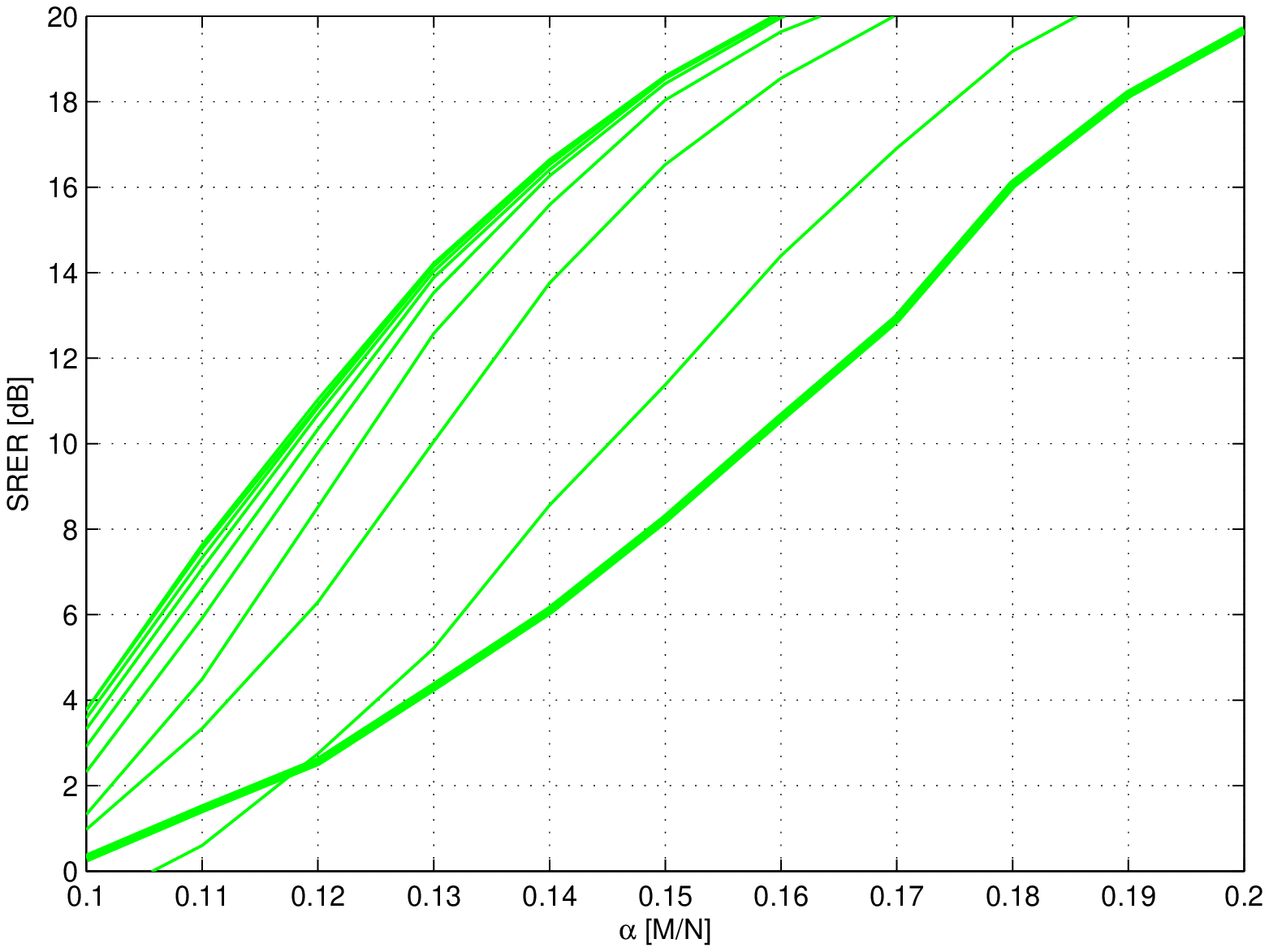}
      \label{fig:domp}
    }
  }
  \subfloat[$\disp$]{
    \resizebox{0.66\columnwidth}{!}{
      \includegraphics[width=\columnwidth]{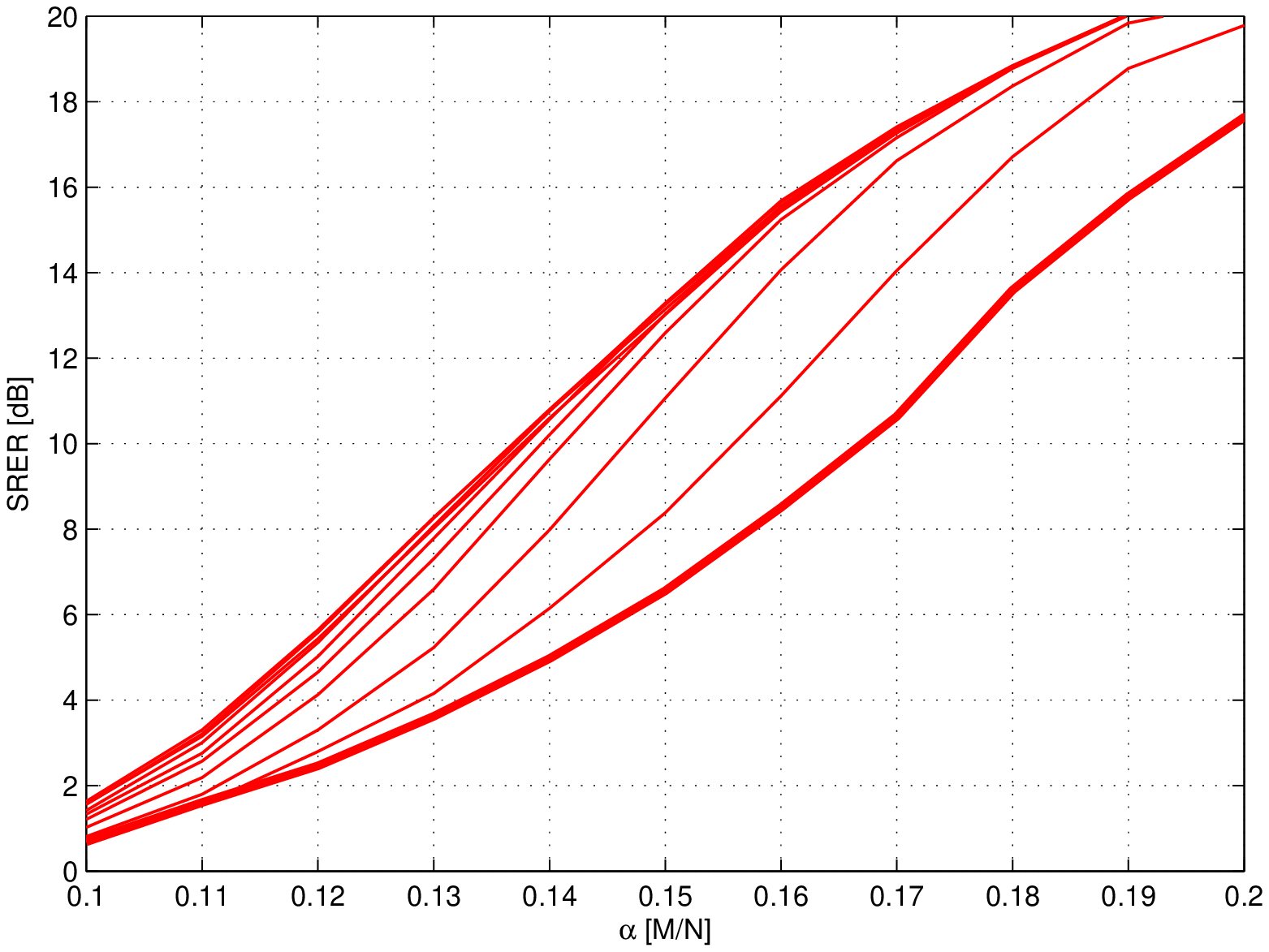}
      \label{fig:sp}
    }
  }
  \subfloat[$\difrogs$]{
    \resizebox{0.66\columnwidth}{!}{
      \includegraphics[width=\columnwidth]{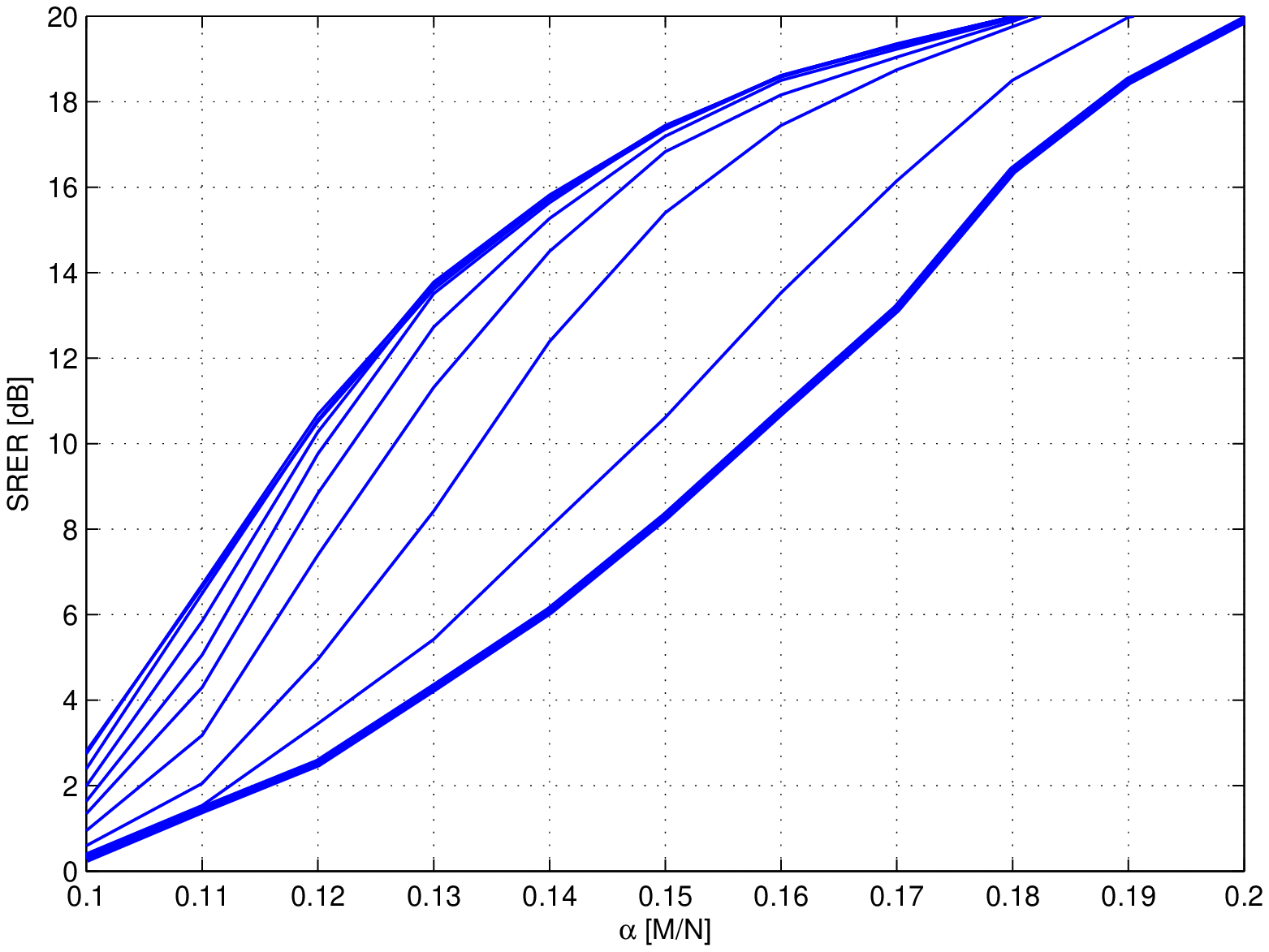}
      \label{fig:frogs}
    }
  }

  \caption{Performance of distributed $\gp$ algorithms for varying network connectivity: \srer versus fraction of measurements $\alpha$ at \smnr = 20 dB. The network connectivity follows $\mathbf{C}_0$, $\mathbf{C}_1$, $\mathbf{C}_2$, \dots , $\C_{L-1}$. The lower-most fat curve corresponds to $\mathbf{C}_0$, which means a standard (disconnected) algorithm and the top most curve corresponds to $\C_{L-1}$, that is a joint algorithm (fully connected network).}
  \label{fig:network_sweep}
\end{figure*}

Using representative setups, we performed computer simulations in order to observe the performance of three \digp algorithms: \diomp, \disp and \difrogs. We compare their performance with two extreme cases: 1) with a centralized solution (i.e., a fully connected network where each node is connected with all other nodes and hence the connection matrix is $\C_{L-1}$) where we refer the algorithms as joint \omp (\textsc{jomp}) \cite{scs2011}, joint \sp (\textsc{jsp}) \cite{scs2011} and joint \frogs (\textsc{jfrogs}); and 2) to a fully disconnected setup (the connections matrix is $\C_0$) where standard \omp, \sp and \frogs are executed independently.
In this paper we focus on the development of a \gp framework for distributed \cs and therefore limit ourselves in the gamut of \gp algorithms. We first discuss the reconstruction performance measures and experimental setups, and then report the performance of all the algorithms for clean and noisy measurement cases.

\subsection{Performance measures and experimental setups}
We use two performance measures. For the first performance measure, we use signal-to-reconstruction-error ratio (\srer) defined as
\begin{eqnarray}
\srer = \frac{ \mathcal{E} \{ \| \mathbf{x} \|_{2}^{2} \} }{\mathcal{E} \{ \| \mathbf{x} - \hat{\mathbf{x}} \|_{2}^{2} \} }, 
\end{eqnarray} 
where $\mathcal{E}$ is the expectation taken over all nodes and all realizations. $\hat{\mathbf{x}}$ is the reconstructed signal vector and our objective is to achieve a higher \srer. Note that we drop the subscript $l$ because we are averaging over all sensors $l$. The \srer is nothing but the inverse of normalized mean square error. Our objective is to achieve a higher \srer.

Next we define another performance measure which provides a direct measure of estimating the underlying support set. This is a distortion measure defined by $d(\T,\Th)= 1- \left( | \T \cap \Th | / |\T|  \right)$ \cite{Gastpar_ASCE} and we have recently used it in \cite{Chatterjee12}. Here, $\T$ is the local support-set, that is $\T = \T^{(c)} \cup \T^{(p)}$. Considering a large number of realizations (signal vectors), we can compute the average of $d(\T,\Th)$. We define the average support-set cardinality error (\asce) as follows
\begin{align}
\asce = \mathcal{E}\left\{ d(\T, \hat{\T}) \right\}= 1 - \mathcal{E}\left\{\frac{|\T \cap \hat{\T}|}{|\T|} \right\}. \label{eqn:ASCE}
\end{align}
Note that the \asce has the range $[0, 1]$ and our objective is to achieve a lower \asce. Along-with \srer, the \asce is used as the second performance evaluation measure because the principle objective of most \gp algorithms is to estimate the underlying support set. 

Next we describe the simulation setups. In any \cs setup, all sparse signals are expected to be exactly reconstructed if the number of measurements are more than a certain threshold value. The computational complexity to test this uniform reconstruction ability is exponentially high. Instead, we can rely on empirical testing, where \srer and \asce are computed for random measurement matrix ensemble. 
For a given network topology $\C_{i} (\text{or }\C_{i,\text{rand}}), \,\, i \in [0,L-1]$ and $\alpha$ as defined in \eqref{eqn:alpha}, the steps of testing strategy are listed as follows:
\begin{packed_enum}
\item Given the parameters $N$, $K^{(c)}$ and $\{ K^{(p)}_l \}_{l=1}^{L}$ choose an $\alpha$ (such that $M$ is an integer). We use same $K^{(p)}_l, \,\forall l$. \label{item:L}
\item Randomly generate a set of $M \times N$ sensing matrices $\left\{\mathbf{A}_l\right\}_{l=1}^L$ where the components are drawn independently from an i.i.d. Gaussian source (i.e. $a_{m,n} \sim \mathcal{N}\left( 0, \frac{1}{M} \right)$) and then scale the columns of $\mathbf{A}_l$ to unit-norm. \label{item:sensing_matrix}
% \item Generate support-sets  $\T^{(c)}$ and $\{ \T^{(p)}_l \}_{l=1}^L$ of cardinality $K^{(c)}$ and $\{K^{(p)}_l\}_{l=1}^L$, respectively. The support-sets are uniformly chosen from $\{1,2,...,N\}$.\label{item:support-set}
\item Randomly generate a set of signal vectors $\{\mathbf{x}_l\}_{l=1}^L$ following Section~\ref{sec:mixed_support_signal_model}. The common and private support-sets are chosen uniformly over the set $\{1,2, \ldots, N \}$. The non-zero components of $\mathbf{x}$ are independently drawn by either of the following two methods. 
\begin{enumerate}
\item {The non-zero components are drawn independently from a standard Gaussian source. This type of signal is referred to as Gaussian sparse signal.}
\item {The non-zero components are set to ones. This type of signal is referred to as binary sparse signal.}
\end{enumerate}
Note that the Gaussian sparse signal is of a compressible nature. That means, in the descending order, the sorted amplitudes of a Gaussian sparse signal vector's components decay fast with respect to the sorted indices. This decaying trend corroborates with several natural signals (for example, wavelet coefficients of an image). On the other hand, a binary sparse signal is not compressible in nature, but of special interest for comparative study, since it represents a particularly challenging case for \omp-type of reconstruction strategies \cite{OMPfirst}, \cite{SPfirst}.  \label{item:data}
\item Compute the measurements $\mathbf{y}_l = \mathbf{A}_l\mathbf{x}_l + \mathbf{w}_l, \forall l \in \{1,2,...,L\}$. Here $\mathbf{w}_l \sim \mathcal{N}(\mathbf{0},\sigma_{w,l}^2 \mathbf{I}_M)$.
\item Apply the \cs algorithms on the data $\{\mathbf{y}_l\}_{l=1}^L$ independently. 
% The connection matrix $\mathbf{C}$ is used to determine how to distribute the data in the network. The algorithms are executed synchronized until they stop independently based on the local stopping criteria.
\end{packed_enum}
In the above simulation procedure, for each node $l \in \{1,2,\ldots, L\}$, $Q$ sets of sensing matrices are created. Then for each sensing node, $P$ sets of data vectors are created. 
In total, we will average over $L\cdot Q\cdot P$ data to evaluate the performance.

Considering the measurement noise $\mathbf{w}_l \sim \mathcal{N} \left( \mathbf{0}, \sigma_{w,l}^{2} \mathbf{I}_{M} \right)$, we define the signal-to-measurement-noise-ratio (\smnr) as
\begin{eqnarray}
\smnr = \frac{ \mathcal{E} \{ \| \mathbf{x} \|_{2}^{2} \} }{ \mathcal{E} \{ \| \mathbf{w} \|_{2}^{2} \} } , 
\end{eqnarray} 
where $\mathcal{E} \{ \| \mathbf{w} \|_{2}^{2} \} = \sigma_{w,l}^{2}M$. For noisy measurement case, we report the experimental results at \smnr 20 dB. 

In the presence of a measurement noise, it is impossible to achieve perfect \cs recovery. On the other hand, for the clean measurement case, perfect \cs recovery of a sparse signal is possible if $\alpha$ exceeds a certain threshold. In the spirit of using \cs for practical applications with less number of measurements at clean and noisy conditions, we are mainly interested in a lower range of $\alpha$ where performances of the contesting algorithms can be fairly compared.

\subsection{Experimental Results}
Using $N = 500$, $K^{(c)} = 10$, $K^{(p)} = 10$, $Q = 100$, $P = 100$ and $L = 10$, we performed experiments. That means, we used 500-dimensional sparse signal vectors with sparsity level less that or equal to $K^{(c)} + K^{(p)} = 20$. Such a $4\%$ sparsity level is chosen in accordance with real life scenarios, for example most of the energy of an image signal in the wavelet domain is concentrated within $2-4\%$ coefficients \cite{Candes:an_introduction_to}.
There are $L = 10$ nodes and for each node, we created $Q = 100$ signal vectors and $P = 100$ sensing matrices. Thus, for a chosen $\alpha$, we evaluated performance by averaging $100 \times 100 \times 10 = 100000$ realizations in each data point. We incremented $\alpha$ from a lower limit to a higher limit in a small step-size (with the constraint that corresponding $M$ is an integer for a value of $\alpha$) and reported the results.   

\subsubsection{Impact of network connectivity}

Let us first observe the effect of increasing network connectivity on the performance of all the three \digp algorithms. The simulation results are shown in \figurename~\ref{fig:network_sweep}. In this case we show the results in the range of $\alpha$ from 0.1 to 0.2. We use Gaussian sparse signal and \smnr = 20 dB. The two extreme results are the performances for $\C_{0}$ and $\C_{9}$. Here we mention that $\C_{0}$ denotes the case of using standard \gp algorithms. We show the \srer results for the three \digp algorithms and observe that the performance improves with the increase of network connectivity. We note that the use of a degree-2 network (connection matrix is $\C_{2}$) leads to much better performance than the standard $\C_{0}$ case. For \diomp using the degree-2 network, at $\alpha=0.14$, we achieve more than 6 dB \srer improvement compared to the $\C_{0}$ case (i.e., the \omp performance). The performance shows a saturation trend as the connectivity increases. Considering a trade-off between 
network connectivity (i.e., communication resource) and performance, we use $\C_{2}$ as the default network for further results. We also comment that $\mathbf{C}_2$ may be considered a quite restrictive network (in the sense that it is not well connected), but still its use leads to a significant gain in performance.

\subsubsection{Fixed vs Random network}
\begin{figure*}[t]
  \linespread{0.5}
  \centering
  \subfloat[\srer (in dB) versus fraction of measurements]{
    \resizebox{0.95\columnwidth}{!}{
      \includegraphics[width=\columnwidth]{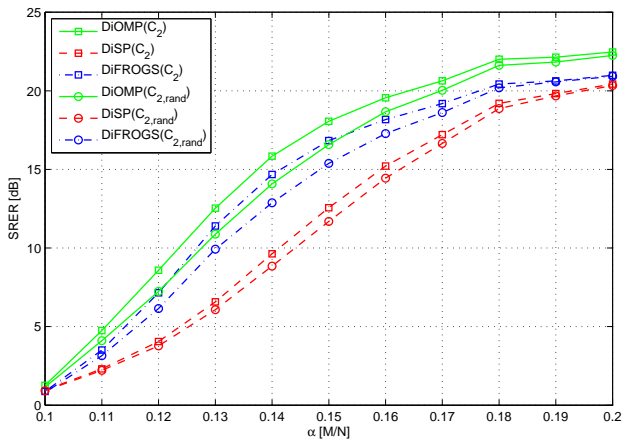}
      \label{fig:rand_a}
    }
  } \qquad
  \subfloat[\asce versus fraction of measurements]{
    \resizebox{0.95\columnwidth}{!}{
      \includegraphics[width=\columnwidth]{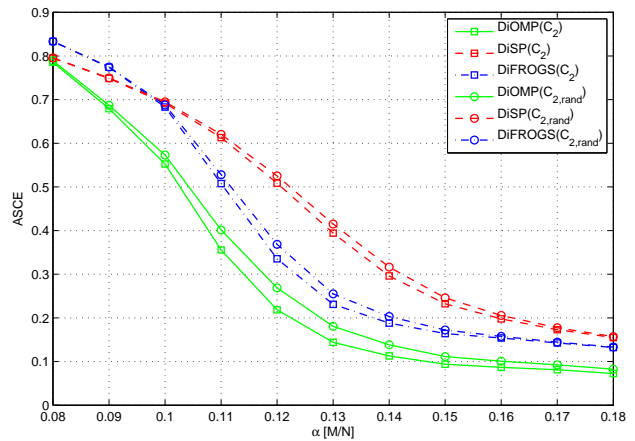}
      \label{fig:rand_b}
    }
  }
  \caption{Comparison of how $\mathbf{C}_2$ compares against $\mathbf{C}_{2,\text{rand}}$ for \digp algorithms with {\textbf{Gaussian}} sparse signal at \textbf{noisy measurement} condition, where $\smnr = 20$ dB. We show performance results against fraction of measurements.}
  \label{fig:rand}
\end{figure*}

The networks considered so far have all been fixed networks represented by $\mathbf{C}_l$ according to the network model described in Section~\ref{sec:network}. However, in the same section we also introduced a random network approach, where each node randomly selects the given number of outgoing neighbors at random referred by $\mathbf{C}_{l,\text{rand}}$. For the network connectivity of two (i.e., $l=2$), we have compared these two sorts of networks in \figurename~\ref{fig:rand}. In \figurename~\ref{fig:rand_a} we show the \srer performance plot and in \figurename~\ref{fig:rand_b} we show the \asce performance. We use Gaussian sparse signal and $\smnr = 20$ dB. The network $\mathbf{C}_{2,\text{rand}}$ was generated a new for each monte-carlo simulation in each point. By studying these figures, we see that the performance for the two different networks is similar. This justifies to use the fixed $\C_l$ for ease of implementation in the controlled experimental simulations.

%Thus, we argue that any approach of the fixed $\mathbf{C}_{l}$ or random $\mathbf{C}_{l,\text{rand}}$ is reasonable.

%it seems the averaged random network $\mathbf{C}_{2,\text{rand}}$ provides for slightly worse performance as compared to the fixed uniform network $\mathbf{C}_2$. However, we see that all algorithms perform about 1-2 dB worse in the same areas so the fixed approach provides for a fair comparison.

\subsubsection{Comparison between algorithms}
Here we provide a comparative study between the three \digp algorithms, the three fully connected joint \gp algorithms and the three completely disconnected, standard, \gp algorithms. So, we compare \diomp, \jomp, \omp, and \disp, \jsp, \sp, and \difrogs, \jfrogs, \frogs algorithms. We use the degree-2 network (connection matrix $\mathbf{C}_2$) for all the \digp algorithms.

\figurename~\ref{fig:bin0smnrinf} shows \srer and \asce results for the case of Gaussian sparse signal at clean measurement conditions. In \figurename~\ref{fig:srerbin0smnrinf}, the three bottom-most \srer curves correspond to the disconnected algorithms. It is important to notice that the new \frogs and \difrogs perform better than \omp and \diomp, respectively. We also note that \sp and \disp perform poorer corresponding to relevant competing algorithms. At $\alpha=0.15$, we note that \diomp provides nearly 15 dB \srer performance improvement compared to the disconnected, standard \omp. Thus, we can comment that our \digp algorithms provide a significant improvement. Similar trends in performance are observed in \figurename~\ref{fig:bin0smnr20} for the noisy measurement condition with \smnr~=~20~dB.

Next we provide the performance results for the binary sparse signal case. \figurename~\ref{fig:bin1smnrinf} shows the results at clean measurement conditions. In this case, the most interesting observation is that the \sp and its allied algorithms (\disp and \jsp) provide significant performance improvements compared to the other relevant competing algorithms. Again we note that \digp algorithms using degree-2 network provide better results than the disconnected stand-alone \gp algorithms. Similar trends in performance are observed in \figurename~\ref{fig:bin1smnr20} for the noisy measurement condition with \smnr = 20 dB. 

Comparing all the results for two different signals at varying measurement conditions and number of measurements, we note that the new \digp algorithms have a promise to provide a good performance. They are capable to provide a good trade-off between network connectivity (i.e., communication resources) and performance.  

\begin{figure*}
{
  \linespread{0.5}
  \centering
  \subfloat[\srer (in dB) versus fraction of measurements]{
    \resizebox{0.95\columnwidth}{!}{
      \includegraphics[width=\columnwidth]{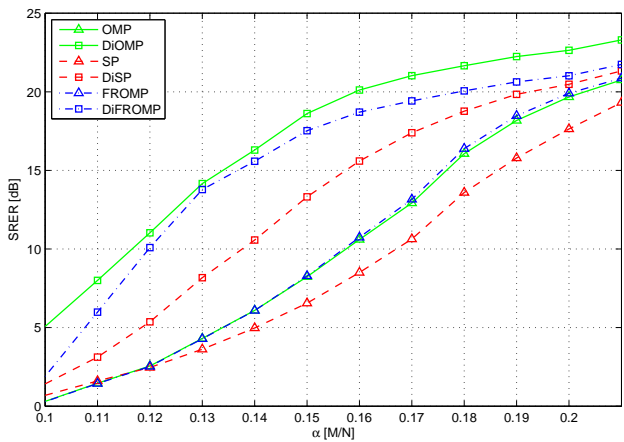}
      \label{fig:practice_a}
    }
  } \qquad
  \subfloat[\asce versus fraction of measurements]{
    \resizebox{0.95\columnwidth}{!}{
      \includegraphics[width=\columnwidth]{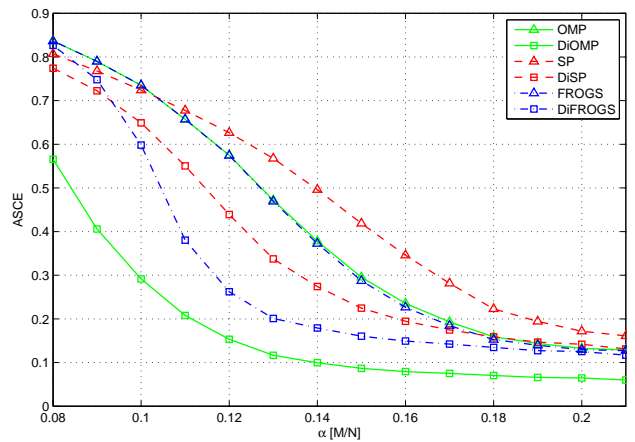}
      \label{fig:practice_b}
    }
  }
  \caption{Simulation results for a network with 100 nodes using the Watts-Strogatz model with connection parameter $p = 3$ and rewiring probability $q = 0.3$. Here we use a \textbf{Gaussian} sparse signal at \textbf{noisy measurement} condition, where $\smnr = 20$ dB.}
  \label{fig:practice}
}
\end{figure*}

{
\subsubsection{Study on a larger network}
We have seen performance curves for fixed and random network setups. However, all these setups are well controlled in order to understand how network connectivity impacts performance. Further, to judge the usage in realistic scenarios, we provide results of a 100-node network in \figurename~\ref{fig:practice}. Here, we use the Watts-Strogatz~\cite{watts1998collective} network model that is claimed to be practically relevant~\cite{aldosari2006topology}. This network model takes two parameters, $q$ and $p$. Using these parameters, first, every node gets connected in a structured way with two-way connections (as opposed to the one-way connections we have previously used) to $q$ neighbors. Then, every connection is rewired with probability $p$ uniformly at random.

In \figurename~\ref{fig:practice}, we have used $q =  3$ and $p = 0.3$ to create one network realization. Using this network realization, each data point is an average over $9\times 10^4$ measurements. The trend of the result is similar with the experiments performed for the controlled, small size networks. For example, at $\alpha = 0.15$, we observe improved performance of $6$ dB higher \srer for \disp compared to \sp, $8$ dB for \difrogs compared to \frogs and $9$ dB for \diomp compared to \omp.
}

\subsubsection{Running-Time Comparison}

At last, we endeavor to provide a running time comparison between the algorithms. This comparison provides a rough idea about the need for computational resources. The running time results are shown in Table~\ref{table:running_time} for varying network connectivity. In this case, we performed simulations for the Gaussian sparse signal case at 20 dB \smnr and $\alpha=0.16$. An interesting point to notice is that even though \frogs is more complex than the \omp, \difrogs requires less computational resource than \diomp. The reason is that \frogs is characterized by \rsupport construction mechanism and hence its use for designing the \difrogs algorithm leads to faster convergence. Another observation is that although \diomp and \difrogs both require longer execution times as the connectivity increases, the same does not necessarily hold for \disp. The reason for this is that the underlying \modsp algorithm iterates fewer times when it is initialized  with a better initial support-set. Thus the increased overhead 
in the voting is balanced by decreased underlying algorithm execution time.

\begin{table}[t]\renewcommand{\arraystretch}{1.2}
\caption{Running time comparison between several \gp and \digp algorithms at varying network connectivity for the typical simulation setup. Times normalized to \sp.}
\label{table:running_time}
\centering
\setlength{\tabcolsep}{2pt}
\begin{tabular}{|c||c|c|c|c|c|c|}
\hline
Network & \multicolumn{3}{c}{\gp algorithms} & \multicolumn{3}{|c|}{\digp algorithms} \\ \cline{2-7}
Connectivity & \omp & \sp & \frogs & \diomp & \disp & \difrogs \\ \hline \hline
$\C_{0}$ & 1.1666 & 1 & 2.5839 & $\times$ & $\times$ & $\times$  \\
$\C_{2}$ & $\times$ & $\times$ & $\times$ & 10.752 & 4.1834 & 7.7554  \\ 
$\C_{3}$ & $\times$ & $\times$ & $\times$ & 10.781 & 4.3961 & 8.1107  \\
$\C_{9}$ & $\times$ & $\times$ & $\times$ & 10.873 & 3.7769 & 8.4220  \\ \hline
\end{tabular}
\end{table}

\textit{Reproducible results:} In the spirit of reproducible results, we provide a package with all necessary \textsc{Matlab} codes in the following website: https://sites.google.com/site/saikatchatt/softwares/. In this package consult the \textsc{readme.txt} file to obtain instructions on how to reproduce the figures presented in this paper.

\section{Conclusion}
For a distributed \cs setup, we have developed a framework for constructing distributed greedy pursuit (\digp) algorithms. Using this framework, we have shown how the two well known greedy algorithms -~\omp and \sp~- can be used for developing two new \digp algorithms. Furthermore, we have created a new \gp algorithm called \frogs using insights gained from a categorization of existing \gp algorithms. Then, based on \frogs we have created a third \digp algorithm. 

In particular we notice that within the new framework, many other \gp algorithms could be used with small modifications as a base in designing new \digp algorithms. Through experimental evaluations we conclude that the new \digp algorithms provide a significant improvement in performance compared to standard, disconnected, \gp algorithms. We also note that the algorithms are capable of providing a trade-off between performance and network connectivity (or a trade-off between performance and communication resource). {Providing analytical performance quantification for distributed \gp algorithms remains as a future research tasks.}

% or


\appendix
\subsection{Modified \omp}\label{app:modOMP}
In this section, we describe the modified \omp (\modomp) algorithm.  Algorithm~\ref{alg:OMP} shows the \modomp.
Instead of initializing with an empty support-set and begin iterating from the first component as in the standard \omp, we allow the \modomp to use an initial support-set as input and continue building the final support-set. This modification reduces to the standard \omp (as shown in~\cite{OMPfirst}) when the initial support-set $\T_{\text{ini}} = \emptyset$.
\begin{algorithm}[ht!]
\caption{: \modomp}\label{alg:OMP}
\mbox{Input: $\mathbf{A}$, $K_{\max}$, $\mathbf{y}$, $\T_{\text{ini}}$.} \newline
\mbox{Initialization: }
\begin{algorithmic}[1]
\STATE $\T_0 \leftarrow \T_{\text{ini}}$
\STATE $\mathbf{r}_0 \leftarrow \resid(\mathbf{y},\mathbf{A}_{\T_0})$
\STATE $k \leftarrow |\T_0|$
\end{algorithmic}
\mbox{Iteration: }
\begin{algorithmic}[1]
\REPEAT
\STATE $k \leftarrow k + 1$
\STATE $\tau_{\max} \leftarrow \maxi\left(\mathbf{A}^T \mathbf{r}_{k-1}, 1\right)$ \label{OMP:selmax}
\STATE $\T_{k} \leftarrow \T_{k-1} \cup \tau_{\max}$ \label{OMP:union}
%\STATE $\hat{\mathbf{x}}_{\T_{k}} \leftarrow \mathbf{A}_{\T_{k}}^{\dagger} \mathbf{y}$ \label{OMP:x}
\STATE $\mathbf{r}_{k} \leftarrow \resid(\mathbf{y}, \mathbf{A}_{\T_{k}})$
\UNTIL{$(k = K_{\max})$}
\end{algorithmic}
\mbox{Output:} 
\begin{algorithmic}[1]
\STATE $\hat{\T} \leftarrow \T_k$
\STATE $\hat{\mathbf{x}}$ ~ such that ~ $\hat{\mathbf{x}}_{\T_k} = \mathbf{A}_{\T_{k}}^{\dagger} \mathbf{y}$ and $\hat{\mathbf{x}}_{\bar{\T}_k} = \mathbf{0}$
\STATE $\eta \leftarrow \|\mathbf{r}_k\|$
\end{algorithmic}
\hrule
\mbox{Functional form:} $(\hat{\T}, \hat{\mathbf{x}}, \eta) \leftarrow \modomp(\mathbf{A}, K_{\max}, \mathbf{y},\T_{\text{ini}}) $
\end{algorithm}
In step~2 of the \emph{initialization}, \modomp finds a residual, where $\T_0 = \T_{\text{ini}}$. If the initial support-set $\T_{\text{ini}} = \emptyset$, the matrix $\mathbf{A}_{\T_0}$ is empty and the residual becomes $\mathbf{y}$. At the $k$'th \emph{iteration} stage \modomp algorithm forms the matched filter, identifies the index corresponding to the largest amplitude (step~3) and adds this to the support-set (step~4). It proceeds with solving a least squares problem with the selected indices (step~5), subtracts the least squares fit and produces a new residual (step~6). This process is updated until $K_{\max}$ components have been picked in the support-set. 
In addition to the support-set estimate $\hat{\T}$, we also output the sparse signal estimate $\hat{\mathbf{x}}$ and the final residual norm $\eta$.

\subsection{Modified \sp}\label{app:modSP}
In this section, we describe the modified \sp (\modsp) in Algorithm~\ref{alg:SP}. Similarly to \modomp, we provide an initial support-set $\T_{\text{ini}}$ to the \modsp. Then, \modsp will continue to improve this support-set building the final support-set. When $\T_{\text{ini}} = \emptyset$, the \modsp reduces to the standard \sp (as shown in~\cite{SPfirst}).
% As mentioned in the description of \disp, we here describe the modified \sp in . The initialization phase has, compared to the regular \sp, been modified so that it can use an initial support-set. The modified \sp reduces to the regular \sp as defined by Dai and Milenkovic . The sub-index $l$ which describes what node a particular variable belongs to is not present in this algorithm because the modified \sp has no knowledge about any networks.
\begin{algorithm}[ht!]
\caption{: \modsp}\label{alg:SP}
\mbox{Input: $\mathbf{A}$, $K_{\max}$, $\mathbf{y}$, $\T_{\text{ini}}$} \newline
\mbox{Initialization: }
\begin{algorithmic}[1]
\STATE $\T' \leftarrow \maxi\left(\mathbf{A}^T \mathbf{y}, K_{\max}\right) \cup \T_{\text{ini}}$
\STATE $\hat{\mathbf{x}}$ ~ such that ~ $\hat{\mathbf{x}}_{\T'} = \mathbf{A}_{\T'}^{\dagger} \mathbf{y}$ and $\hat{\mathbf{x}}_{\bar{\T'}} = \mathbf{0}$ \label{SP:2}
\STATE $\T_{0} \leftarrow \maxi(\hat{\mathbf{x}}, K_{\max})$ \label{SP:3}
\STATE $\mathbf{r}_0 \leftarrow \resid(\mathbf{y}, \mathbf{A}_{\T_0})$
\STATE $k \leftarrow 0$
\end{algorithmic}
\mbox{Iteration:}
\begin{algorithmic}[1]
\REPEAT
\STATE $k \leftarrow k + 1$
\STATE $\T' \leftarrow \maxi\left(\mathbf{A}^T \mathbf{r}_{k-1}, K_{\max}\right) \cup \T_{k-1}$
\STATE $\hat{\mathbf{x}}$ ~ such that ~ $\hat{\mathbf{x}}_{\T'} = \mathbf{A}_{\T'}^{\dagger} \mathbf{y}$ and $\hat{\mathbf{x}}_{\bar{\T'}} = \mathbf{0}$
\STATE $\T_{k} \leftarrow \maxi(\hat{\mathbf{x}}, K_{\max})$
\STATE $\mathbf{r}_k \leftarrow \resid(\mathbf{y}, \mathbf{A}_{\T_k})$
\UNTIL{$(\|\mathbf{r}_k\| \geq \|\mathbf{r}_{k-1}\|)$}
\STATE $k \leftarrow k - 1$ \hfill (`Previous iteration count')
\end{algorithmic}
\mbox{Output: }
\begin{algorithmic}[1]
\STATE $\hat{\T} \leftarrow \T_k$
\STATE $\hat{\mathbf{x}}$ ~ such that ~ $\hat{\mathbf{x}}_{\T_k} = \mathbf{A}_{\T_{k}}^{\dagger} \mathbf{y}$ and $\hat{\mathbf{x}}_{\bar{\T}_k} = \mathbf{0}$
\STATE $\eta \leftarrow \|\mathbf{r}_{k}\|$
\end{algorithmic}
\hrule
\mbox{Functional form:} $(\hat{\T}, \hat{\mathbf{x}}, \eta) \leftarrow \modsp(\mathbf{A}, K_{\max}, \mathbf{y},\T_{\text{ini}}) $
\end{algorithm}
At $k$'th \emph{iteration} stage, the modified \sp algorithm forms the matched filter $\mathbf{A}^T \mathbf{r}_{k-1}$, identifies the $K_{\max}$ most prominent indices and merges them with the old support-set~(step~3). This support-set $\T'$ is likely to have a cardinality larger than $K_{\max}$ (usually $K_{\max} \leq |\T'| \leq 2 K_{\max}$). 
The algorithm then forms a least squares estimate with the selected indices of $\T'$ and identifies the indices corresponding to the $K_{\max}$ largest amplitude~(step~4 and 5). The \modsp then finds the residual~(step~6) and repeats the iteration process until the residual norm does not increase. 
In addition to the support-set estimate $\hat{\T}$, we also output the sparse signal estimate $\hat{\x}$ and the final residual norm $\eta$.

{
\subsubsection{Proof of Proposition~\ref{prop:modsp}}
To prove Proposition~\ref{prop:modsp}, we first need to get acquainted with the restricted isometry property (RIP).
\begin{definition}[RIP: Restricted Isometry Property]\label{def:rip}
A matrix $\A$ satisfies the RIP with Restricted Isometry Constant (RIC) $\delta_K$ if
\begin{align}
  (1-\delta_K)\|\x\|^2 \leq \| \A \x \|^2 \leq (1+\delta_K) \|\x\|^2,
\end{align}
for all $K$-sparse vectors $\x$ where $0 \leq \delta_K < 1$.
\end{definition}
Based on RIP, several \gp algorithms shows a recurrence, which is a performance relation between the $k$'th and $k-1$'th iteration.
  \begin{proposition}[Theorem 2.2 in~\cite{giryes:sptrans}, Theorem 3.1 in~\cite{giel2010tr}]~\label{prop:recurrence}
    For a $K$-sparse vector $\x$, where $\delta_{3K} \leq 0.139$, the \sp and \modsp solution at the $k$th iteration satisfies the recurrence
    \begin{align}
      \| \x_{\bar{\T}_k} \|_2 \leq  0.5 \|\x_{\bar{\T}_{k-1}} \|_2 + 8.22 \| \A^T_{\T_{\w}} \w \|_2. \label{eqn:recurrence}
    \end{align}
  \end{proposition}
  Using Proposition~\ref{prop:recurrence}, we now continue to the main proof of this section.
  \begin{IEEEproof}[Proof of Proposition~\ref{prop:modsp}]
    To find the solution, we apply the recurrence inequality \eqref{eqn:recurrence} recursively
    \begin{align}
      \| \x_{\bar{\T}_k} \|_2 & \leq  0.5 \|\x_{\bar{\T}_{k-1}} \|_2 + 8.22 \| \A^T_{\T_{\w}} \w \|_2 \\
      & \leq  0.5^2 \|\x_{\bar{\T}_{k-2}} \|_2 + 8.22 (0.5+1)  \| \A^T_{\T_{\w}} \w \|_2 \\
      & \dots \\
      & \leq 0.5^{k^*} \|\x_{\bar{\T}_{k-k^*}} \|_2 + 8.22 \sum_{i=0}^{k^*-1}{0.5^i}  \| \A^T_{\T_{\w}} \w \|_2 \\
      & \leq 2^{-k^*} \|\x_{\bar{\T}_{k-k^*}} \|_2 + 16.44 \| \A^T_{\T_{\w}} \w \|_2. \label{eqn:modsp_end}
    \end{align}
    We now let the number of iterations be $k = k_{\modsp}^* = \left\lceil \log_2\left( \frac{\| \x_{\bar{\T}_{\text{ini}}} \|_2}{\| \A^T_{\T_{\w}} \w \|_2} \right)\right\rceil$, where we have that $\|\x_{\bar{\T}_{k-k^*}} \|_2 = \|\x_{\bar{\T}_{0}} \|_2 = \|\x_{\bar{\T}_{\text{ini}}} \|_2$. Plugging $k_{\modsp}^*$ into \eqref{eqn:modsp_end} we get that
  \begin{align}
    \| \x_{\bar{\T}_{k^*}} \|_2 & \leq 2^{-k^*} \|\x_{\bar{\T}_{k-k^*}} \|_2 + 16.44 \| \A^T_{\T_{\w}} \w \|_2 \\
    & = (1 + 16.44) \| \A^T_{\T_{\w}} \w \|_2. \label{eqn:modsp_final_bound}
  \end{align}
  We now introduce the following inequality (derived in e.g. Lemma 3 in~\cite{SPfirst} but also applicable here)
  \begin{align}
    \| \x - \hat{\x}_{\modsp} \| \leq \frac{1}{1-\delta_{3K}} \| \x_{\bar{\T}_{k^*}}  \| + \frac{1}{1-\delta_{3K}} \| \A^T_{\T_{\w}} \w  \|, \label{eqn:bound}
  \end{align}
  where $\hat{\x}_{\modsp}$ is constructed such that $\hat{\mathbf{x}}_{\T_{k^*}} = \mathbf{A}_{\T_{k^*}}^{\dagger} \mathbf{y}$ and $\hat{\mathbf{x}}_{\bar{\T}_{k^*}} = \mathbf{0}$
 Now, applying \eqref{eqn:bound} to \eqref{eqn:modsp_final_bound} gives us
  \begin{align}
    \| \x - \xh_{\modsp} \|_2 \leq 21.41 \| \A^T_{\T_{\w}} \w \|_2.    
  \end{align}
\end{IEEEproof}
}

%\section*{Acknowledgment}

\ifCLASSOPTIONcaptionsoff
  \newpage
\fi



\bibliographystyle{IEEEtran}
% argument is your BibTeX string definitions and bibliography database(Salamis)
\bibliography{../references/cs.bib}

\begin{thebibliography}{10}
\providecommand{\url}[1]{#1}
\csname url@samestyle\endcsname
\providecommand{\newblock}{\relax}
\providecommand{\bibinfo}[2]{#2}
\providecommand{\BIBentrySTDinterwordspacing}{\spaceskip=0pt\relax}
\providecommand{\BIBentryALTinterwordstretchfactor}{4}
\providecommand{\BIBentryALTinterwordspacing}{\spaceskip=\fontdimen2\font plus
\BIBentryALTinterwordstretchfactor\fontdimen3\font minus
  \fontdimen4\font\relax}
\providecommand{\BIBforeignlanguage}[2]{{%
\expandafter\ifx\csname l@#1\endcsname\relax
\typeout{** WARNING: IEEEtran.bst: No hyphenation pattern has been}%
\typeout{** loaded for the language `#1'. Using the pattern for}%
\typeout{** the default language instead.}%
\else
\language=\csname l@#1\endcsname
\fi
#2}}
\providecommand{\BIBdecl}{\relax}
\BIBdecl

\bibitem{CS:donoho}
D.~Donoho, ``Compressed sensing,'' \emph{IEEE Trans. Inf. Theory}, vol.~52,
  no.~4, pp. 1289 --1306, April 2006.

\bibitem{CRT2006}
E.~Candes, J.~Romberg, and T.~Tao, ``Robust uncertainty principles: exact
  signal reconstruction from highly incomplete frequency information,''
  \emph{IEEE Trans. Inf. Theory}, vol.~52, no.~2, pp. 489 -- 509, Feb. 2006.

\bibitem{gabs2010}
A.~Yang, M.~Gastpar, R.~Bajcsy, and S.~Sastry, ``Distributed sensor perception
  via sparse representation,'' \emph{Proc. of the IEEE}, vol.~98, no.~6, pp.
  1077 --1088, June 2010.

\bibitem{DistPSD2}
F.~Zeng, C.~Li, and Z.~Tian, ``Distributed compressive spectrum sensing in
  cooperative multihop cognitive networks,'' \emph{IEEE Journal Selected Topics
  in Signal Processing}, vol.~PP, no.~99, pp. 1 --1, 2010.

\bibitem{DistPSD3}
J.~Bazerque and G.~Giannakis, ``Distributed spectrum sensing for cognitive
  radio networks by exploiting sparsity,'' \emph{IEEE Trans. Signal
  Processing}, vol.~58, no.~3, pp. 1847 --1862, Mar. 2010.

\bibitem{Distributed1}
Q.~Ling and Z.~Tian, ``Decentralized support detection of multiple measurement
  vectors with joint sparsity,'' in \emph{Proc.~IEEE Int.~Conf.~Acoustics,
  Speech and Signal Processing (ICASSP)}, may 2011, pp. 2996 --2999.

\bibitem{scs2011}
D.~Sundman, S.~Chatterjee, and M.~Skoglund, ``Greedy pursuits for compressed
  sensing of jointly sparse signals,'' in \emph{Proc.~Eur. Sig. Proc. Conf.},
  Aug 2011.

\bibitem{DistCompSens}
D.~Baron, M.~F. Duarte, M.~B. Wakin, S.~Sarvotham, and R.~G. Baraniuk,
  ``Distributed compressive sensing,'' http://arxiv.org/abs/0901.3403, 2009.

\bibitem{DistCSjournal1}
M.~Duarte, S.~Sarvotham, D.~Baron, M.~Wakin, and R.~Baraniuk, ``Distributed
  compressed sensing of jointly sparse signals,'' \emph{Proc.~Asilomar
  Conf.~Signals, Sys., and Comp.}, pp. 1537 -- 1541, Oct. 2005.

\bibitem{SSA1}
G.~Davis, S.~Mallat, and M.~Avellaneda, ``{Greedy adaptive approximation},''
  \emph{J. Constr. Approx.}, vol.~13, pp. 57--98, 1997.

\bibitem{SSA2}
D.~Leviatan and V.~Temlyakov, ``Simultaneous approximation by greedy
  algorithms,'' \emph{Advances in Computational Mathematics}, vol.~25, no. 1-3,
  pp. 73--90, Jul. 2006.

\bibitem{Distapp1}
S.~Cotter, B.~Rao, K.~Engan, and K.~Kreutz-Delgado, ``Sparse solutions to
  linear inverse problems with multiple measurement vectors,'' \emph{IEEE
  Trans. Signal Processing}, vol.~53, no.~7, pp. 2477 -- 2488, Jul. 2005.

\bibitem{MMV1}
J.~Chen and X.~Huo, ``Theoretical results on sparse representations of
  multiple-measurement vectors,'' \emph{IEEE Trans. Signal Processing},
  vol.~54, no.~12, pp. 4634 --4643, dec. 2006.

\bibitem{SOMP}
J.~Tropp, A.~Gilbert, and M.~Strauss, ``Simultaneous sparse approximation via
  greedy pursuit,'' in \emph{Proc.~IEEE Int.~Conf.~Acoustics, Speech and Signal
  Processing (ICASSP)}, vol.~5, mar. 2005, pp. v/721 -- v/724 Vol. 5.

\bibitem{Rakotomamonjy20111505}
\BIBentryALTinterwordspacing
A.~Rakotomamonjy, ``Surveying and comparing simultaneous sparse approximation
  (or group-lasso) algorithms,'' \emph{Signal Processing}, vol.~91, no.~7, pp.
  1505 -- 1526, 2011. [Online]. Available:
  \url{http://www.sciencedirect.com/science/article/pii/S0165168411000272}
\BIBentrySTDinterwordspacing

\bibitem{Mota:distributed_basis_pursuit}
J.~Mota, J.~Xavier, P.~Aguiar, and M.~Puschel, ``Distributed basis pursuit,''
  \emph{IEEE Trans. Signal Processing}, vol.~PP, no.~99, p.~1, 2011.

\bibitem{Qing:decentralized}
L.~Qing, W.~Zaiwen, and Y.~Wotao, ``Decentralized jointly sparse optimization
  by reweighted lq minimization,'' \emph{IEEE Trans. Signal Processing},
  vol.~61, no.~5, pp. 1165--1170, 2013.

\bibitem{convopt}
S.~Boyd and L.~Vandenberghe, \emph{Convex Optimization}.\hskip 1em plus 0.5em
  minus 0.4em\relax New York, NY, USA: Cambridge University Press, 2004.

\bibitem{Chatterjee12}
S.~Chatterjee, D.~Sundman, M.~Vehkaper\"{a}, and M.~Skoglund,
  ``Projection-based and look ahead strategies for atom selection,'' \emph{IEEE
  Trans. Signal Processing}, vol.~60, no.~2, pp. 634 --647, feb 2012.

\bibitem{algo_sim_sparse}
J.~Tropp, A.~Gilbert, and M.~Strauss, ``{Algorithms for simultaneous sparse
  approximation. Part I: Greedy pursuit},'' \emph{Signal Processing}, vol.~86,
  no.~3, pp. 572--588, Mar. 2006.

\bibitem{Sundman:a_greedy_pursuit_algorithm}
D.~Sundman, S.~Chatterjee, and M.~Skoglund, ``A greedy pursuit algorithm for
  distributed compressed sensing,'' in \emph{Proc.~IEEE Int.~Conf.~Acoustics,
  Speech and Signal Processing (ICASSP)}, Kyoto, Japan, Mar. 2012, pp.
  2729--2732.

\bibitem{wimalajeewa:cooperative}
T.~Wimalajeewa and P.~Varshney, ``Cooperative sparsity pattern recovery in
  distributed networks via distributed-omp,'' in \emph{Proc.~IEEE
  Int.~Conf.~Acoustics, Speech and Signal Processing (ICASSP)}, Vancouver,
  Canada, May 2013.

\bibitem{grsv2008}
R.~{G}ribonval, H.~{R}auhut, K.~{S}chnass, and P.~{V}andergheynst, ``{A}toms of
  all channels, unite! {A}verage case analysis of multi-channel sparse recovery
  using greedy algorithms,'' \emph{J. Fourier Anal. and Appl.}, vol.~14, no.~5,
  pp. 655--687, 2008.

\bibitem{OMPfirst}
J.~Tropp and A.~Gilbert, ``Signal recovery from random measurements via
  orthogonal matching pursuit,'' \emph{IEEE Trans. Inf. Theory}, vol.~53,
  no.~12, pp. 4655 --4666, Dec. 2007.

\bibitem{SPfirst}
W.~Dai and O.~Milenkovic, ``Subspace pursuit for compressive sensing signal
  reconstruction,'' \emph{IEEE Trans. Inf. Theory}, vol.~55, no.~5, pp. 2230
  --2249, May 2009.

\bibitem{Sundman:diprsp}
D.~Sundman, D.~Zachariah, and S.~Chatterjee, ``Distributed predictive subspace
  pursuit,'' in \emph{Proc.~IEEE Int.~Conf.~Acoustics, Speech and Signal
  Processing (ICASSP)}, Vancouver, Canada, May 2013.

\bibitem{sucs2010}
D.~Sundman, S.~Chatterjee, and M.~Skoglund, ``On the use of compressive
  sampling for wide-band spectrum sensing,'' in \emph{Proc.~IEEE Int. Symp.
  Signal Processing and Inf. Tech. (ISSPIT)}, Dec. 2010, pp. 354 --359.

\bibitem{Kirmani:codac}
A.~Kirmani, A.~Colaco, F.~Wong, and V.~Goyal, ``{C}o{DAC}: A compressive depth
  acquisition camera framework,'' in \emph{Proc.~IEEE Int.~Conf.~Acoustics,
  Speech and Signal Processing (ICASSP)}, Kyoto, Japan, Mar. 2012, pp.
  5425--5428.

\bibitem{Wu:spherical_microphone}
P.~Wu, N.~Epain, and C.~Jin, ``A dereverberation algorithm for spherical
  microphone arrays using compressed sensing techniques,'' in \emph{Proc.~IEEE
  Int.~Conf.~Acoustics, Speech and Signal Processing (ICASSP)}, Kyoto, Japan,
  Mar. 2012, pp. 4053--4056.

\bibitem{watts1998collective}
D.~J. Watts and S.~H. Strogatz, ``Collective dynamics of 'small-world'
  networks,'' \emph{Nature}, vol. 393, pp. 440--442, 1998.

\bibitem{Pattersson:distributed}
S.~Patterson, Y.~Eldar, and I.~Keidar, ``Distributed sparse signal recovery for
  sensor networks,'' in \emph{Proc.~IEEE Int.~Conf.~Acoustics, Speech and
  Signal Processing (ICASSP)}, Vancouver, Canada, May 2013.

\bibitem{Boyd:2011:DOS:2185815.2185816}
\BIBentryALTinterwordspacing
S.~Boyd, N.~Parikhl, E.~Chu, B.~Peleato, and J.~Eckstein, ``Distributed
  optimization and statistical learning via the alternating direction method of
  multipliers,'' \emph{Found. Trends Mach. Learn.}, vol.~3, no.~1, pp. 1--122,
  Jan. 2011. [Online]. Available: \url{http://dx.doi.org/10.1561/2200000016}
\BIBentrySTDinterwordspacing

\bibitem{dantzig1965linear}
\BIBentryALTinterwordspacing
G.~Dantzig, \emph{Linear Programming and Extensions}, ser. Landmarks in Physics
  and Mathematics.\hskip 1em plus 0.5em minus 0.4em\relax Princeton University
  Press, 1965. [Online]. Available:
  \url{http://books.google.se/books?id=2j46uCX5ZAYC}
\BIBentrySTDinterwordspacing

\bibitem{Dutta:datamining}
H.~Dutta and H.~Kargupta, ``Distributed linear programming and resource
  management for data mining in distributed environments,'' in \emph{IEEE
  International Conference on Data Mining Workshops}, 2008, pp. 543--552.

\bibitem{Yarmish:2001:DIS:933505}
G.~Yarmish, ``A distributed implementation of the simplex method,'' Ph.D.
  dissertation, Polytechnic University, Brooklyn, NY, USA, 2001, aAI3006399.

\bibitem{Stunkel:1989:HIS:63047.63104}
\BIBentryALTinterwordspacing
C.~B. Stunkel and D.~A. Reed, ``Hypercube implementation of the simplex
  algorithm,'' in \emph{Proceedings of the third conference on Hypercube
  concurrent computers and applications - Volume 2}, ser. C3P.\hskip 1em plus
  0.5em minus 0.4em\relax New York, NY, USA: ACM, 1988, pp. 1473--1482.
  [Online]. Available: \url{http://doi.acm.org/10.1145/63047.63104}
\BIBentrySTDinterwordspacing

\bibitem{blumensath2009iterative}
T.~Blumensath and M.~E. Davies, ``Iterative hard thresholding for compressed
  sensing,'' \emph{Applied and Computational Harmonic Analysis}, vol.~27,
  no.~3, pp. 265--274, 2009.

\bibitem{zachariah:dip}
D.~Zachariah, S.~Chatterjee, and M.~Jansson, ``Dynamic iterative pursuit,''
  \emph{IEEE Trans. Signal Processing}, vol.~60, no.~9, pp. 4967--4972, 2012.

\bibitem{netro2009}
\BIBentryALTinterwordspacing
D.~Needell and J.~A. Tropp, ``{CoSaMP: Iterative signal recovery from
  incomplete and inaccurate samples☆},'' \emph{Appl. and Comp. Harm.
  Analysis}, vol.~26, pp. 301--321, May 2009. [Online]. Available:
  \url{http://dx.doi.org/10.1016/j.acha.2008.07.002}
\BIBentrySTDinterwordspacing

\bibitem{donoho2006}
D.~L. Donoho, Y.~Tsaig, I.~Drori, and J.~luc Starck, ``Sparse solution of
  underdetermined linear equations by stagewise orthogonal matching pursuit,''
  Tech. Rep., 2006.

\bibitem{backtrackingomp}
H.~Huang and A.~Makur, ``Backtracking-based matching pursuit method for sparse
  signal reconstruction,'' \emph{IEEE Signal Processing Letters}, vol.~18,
  no.~7, pp. 391 --394, july 2011.

\bibitem{swectwsuch}
D.~Sundman, S.~Chatterjee, and M.~Skoglund, ``Look ahead parallel pursuit,'' in
  \emph{2011 IEEE Swedish Communication Technologies Workshop (Swe-CTW)}, oct.
  2011, pp. 114 --117.

\bibitem{Needell:signal_recovery_incomplete}
D.~Needell and R.~Vershynin, ``Signal recovery from incomplete and inaccurate
  measurements via regularized orthogonal matching pursuit,'' \emph{IEEE
  Journal of Selected Topics in Signal Processing}, vol.~4, no.~2, pp.
  310--316, Apr. 2010.

\bibitem{CMP_2007}
M.~Christensen and S.~Jensen, ``The cyclic matching pursuit and its application
  to audio modeling and coding,'' in \emph{Proc.~Asilomar Conf.~Signals, Sys.,
  and Comp.}, nov. 2007, pp. 550--554.

\bibitem{CMP_2011}
B.~Sturm, M.~Christensen, and R.~Gribonval, ``Cyclic pure greedy algorithms for
  recovering compressively sampled sparse signals,'' in \emph{Proc.~Asilomar
  Conf.~Signals, Sys., and Comp.}, nov. 2011, pp. 1143 --1147.

\bibitem{giryes:sptrans}
R.~Giryes and M.~Elad, ``Rip-based near-oracle performance guarantees for sp,
  cosamp, and iht,'' \emph{IEEE Trans. Signal Processing}, vol.~60, no.~3, pp.
  1465--1468, 2012.

\bibitem{giel2010tr}
------, ``Rip-based near-oracle performance guarantees for subspace-pursuit,
  cosamp, and iterative hard-thresholding,'' Technion, Tech. Rep., 2010.

\bibitem{Gastpar_ASCE}
G.~Reeves and M.~Gastpar, ``A note on optimal support recovery in compressed
  sensing,'' \emph{Proc.~Asilomar Conf.~Signals, Sys., and Comp.}, pp. 1576
  --1580, nov. 2009.

\bibitem{Candes:an_introduction_to}
E.~Candes and M.~Wakin, ``An introduction to compressive sampling,'' \emph{IEEE
  Signal Processing Magazine}, vol.~25, no.~2, pp. 21 --30, Mar. 2008.

\bibitem{aldosari2006topology}
S.~A. Aldosari and J.~M. Moura, ``Topology of sensor networks in distributed
  detection,'' in \emph{Proc.~IEEE Int.~Conf.~Acoustics, Speech and Signal
  Processing (ICASSP)}, vol.~5.\hskip 1em plus 0.5em minus 0.4em\relax IEEE,
  2006, pp. V--V.

\end{thebibliography}


% \end{thebibliography}

%\begin{biography}[{\includegraphics[width=1in,height=1.25in,clip,keepaspectratio]{mshell}}]{Michael Shell}

% \begin{IEEEbiography}{Dennis Sundman}
% \end{IEEEbiography}
% \begin{IEEEbiographynophoto}{Saikat Chatterjee}

%\newpage

% \begin{IEEEbiographynophoto}{Mikael Skoglund}

\begin{landscape}
\begin{figure}[ht!]
  \linespread{0.5}
  \centering
  \subfloat[\srer (in dB) versus fraction of measurements]{
    \resizebox{0.46\columnwidth}{!}{
      \includegraphics[width=\columnwidth]{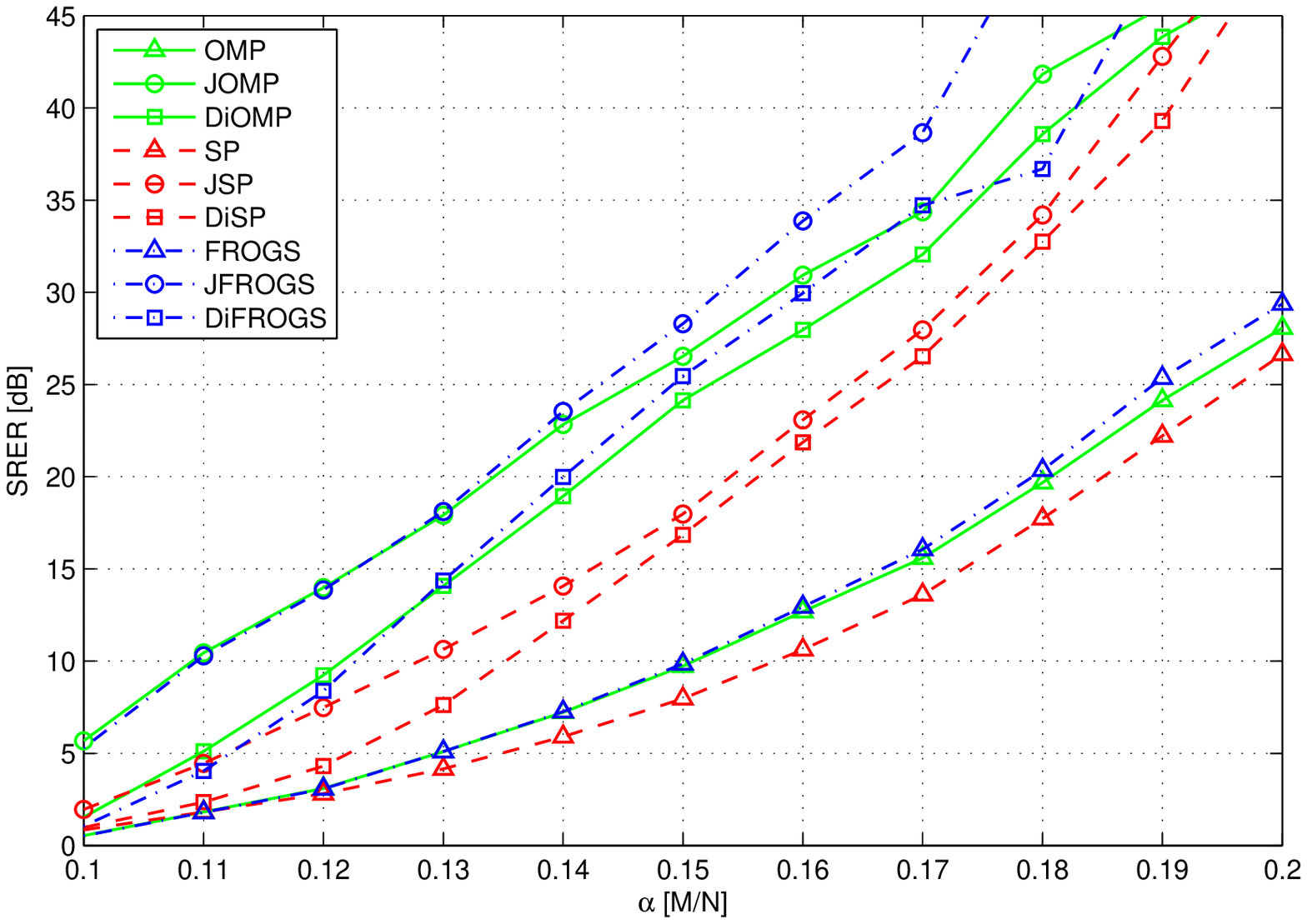}
      \label{fig:srerbin0smnrinf}
    }
  } \qquad
  \subfloat[\asce versus fraction of measurements]{
    \resizebox{0.46\columnwidth}{!}{
      \includegraphics[width=\columnwidth]{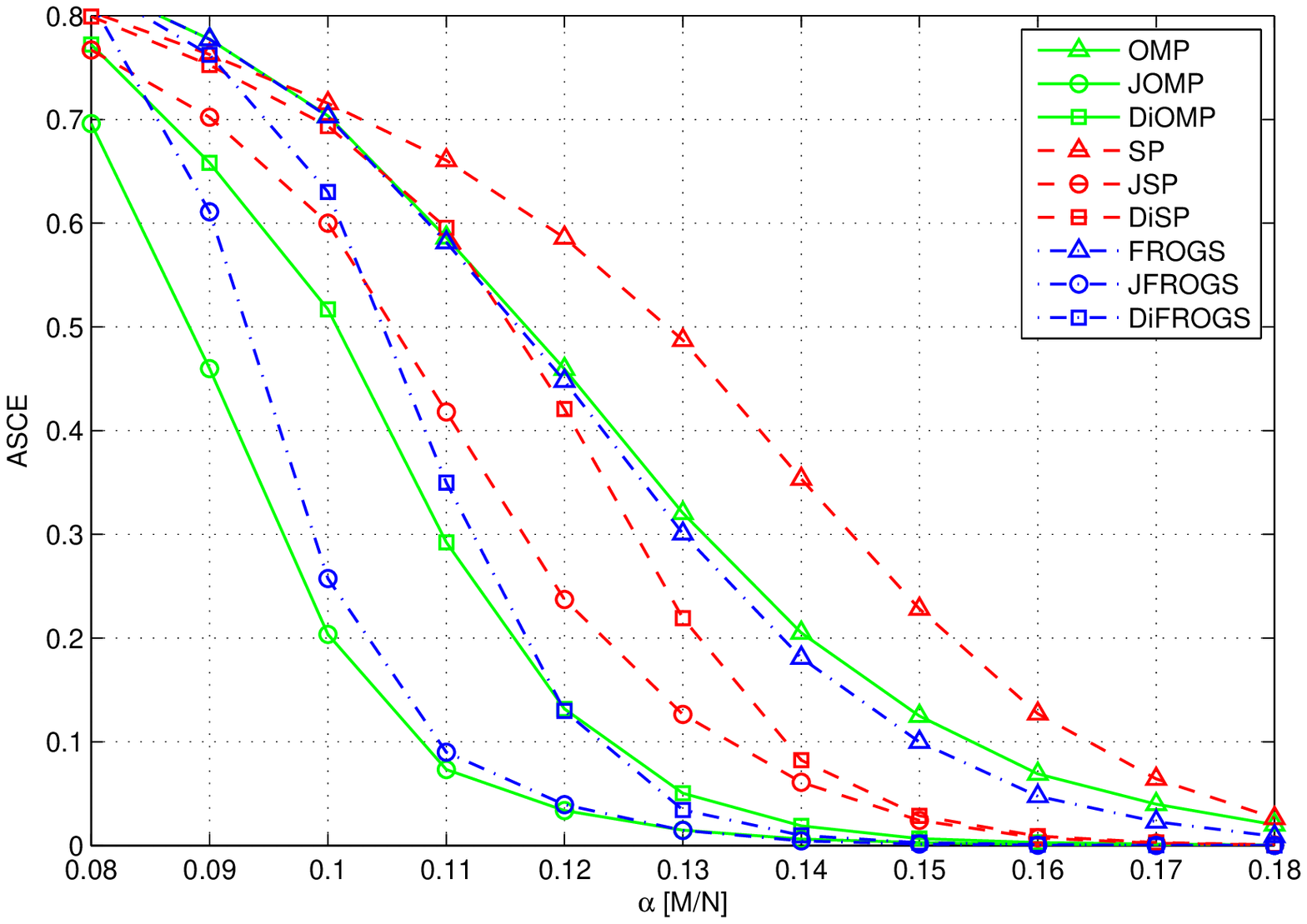}
      \label{fig:ascebin0smnrinf}
    }
  }
  \caption{Comparison of \gp, \digp and joint \gp algorithms for {\textbf{Gaussian}} sparse signal at \textbf{clean measurement} condition. We show performance results against fraction of measurements.}
  \label{fig:bin0smnrinf}
  \subfloat[\srer (in dB) versus fraction of measurements]{
    \resizebox{0.46\columnwidth}{!}{
      \includegraphics[width=\columnwidth]{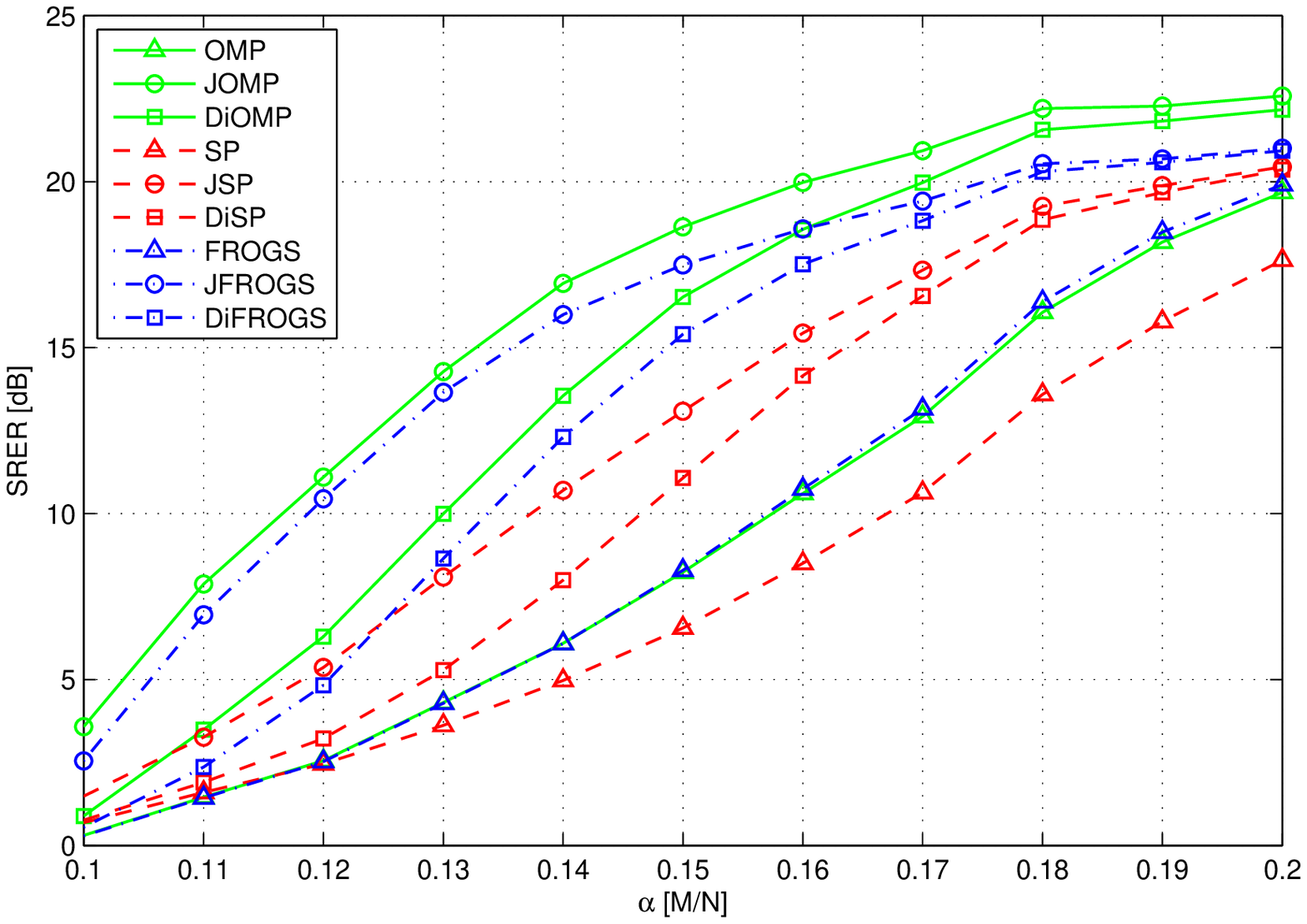}
      \label{fig:srerbin0smnr20}
    }
  } \qquad
  \subfloat[\asce versus fraction of measurements]{
    \resizebox{0.46\columnwidth}{!}{
      \includegraphics[width=\columnwidth]{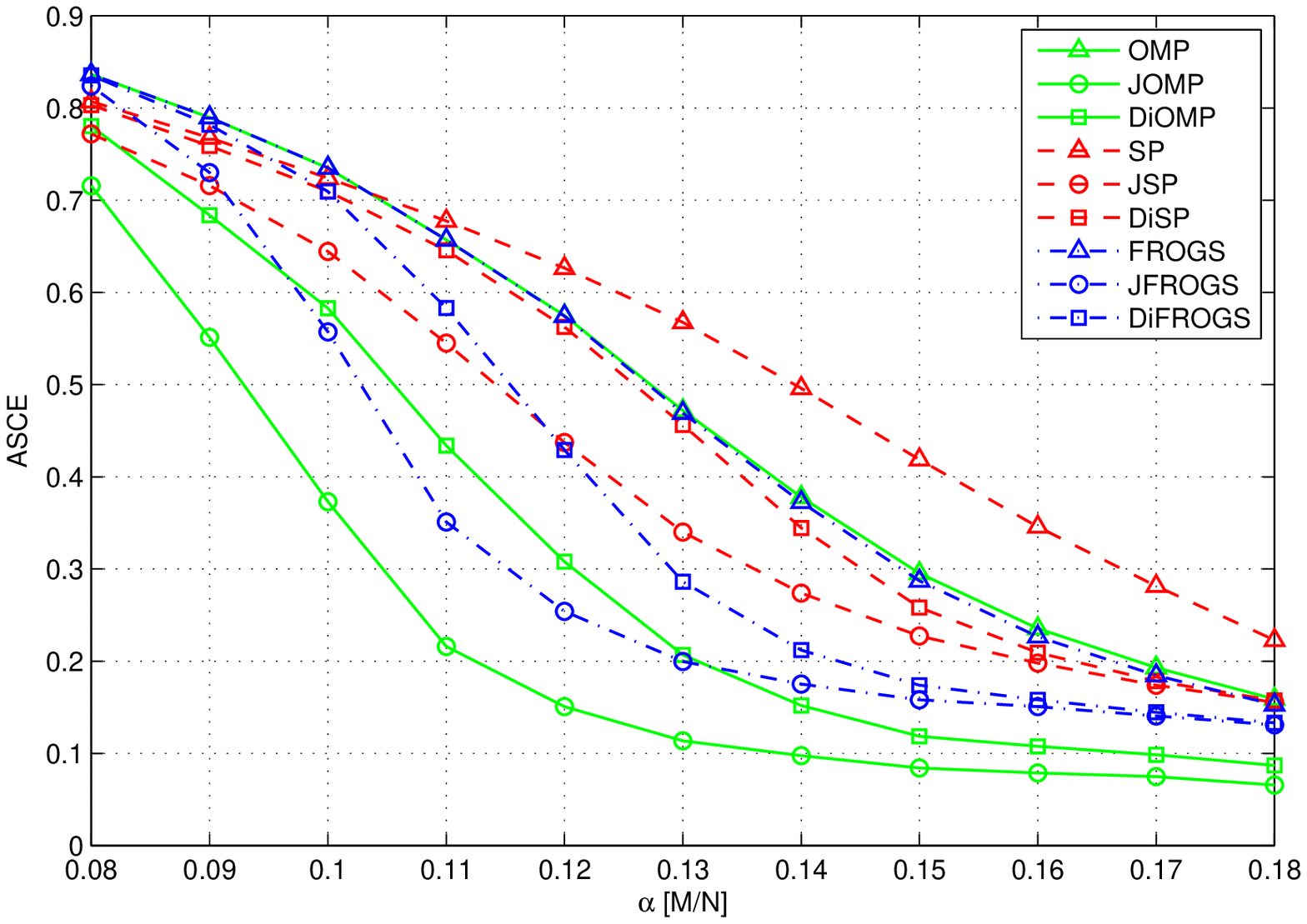}
      \label{fig:ascebin0smnr20}
    }
  }
  \caption{Comparison of \gp, \digp and joint \gp algorithms for {\textbf{Gaussian}} sparse signal at \textbf{noisy measurement} condition, where \smnr = 20 dB. We show performance results against fraction of measurements.}
  \label{fig:bin0smnr20}
\end{figure}
\end{landscape}
\begin{landscape}
\begin{figure}[ht!]
  \linespread{0.5}
  \centering
  \subfloat[\srer (in dB) versus fraction of measurements]{
    \resizebox{0.46\columnwidth}{!}{
      \includegraphics[width=\columnwidth]{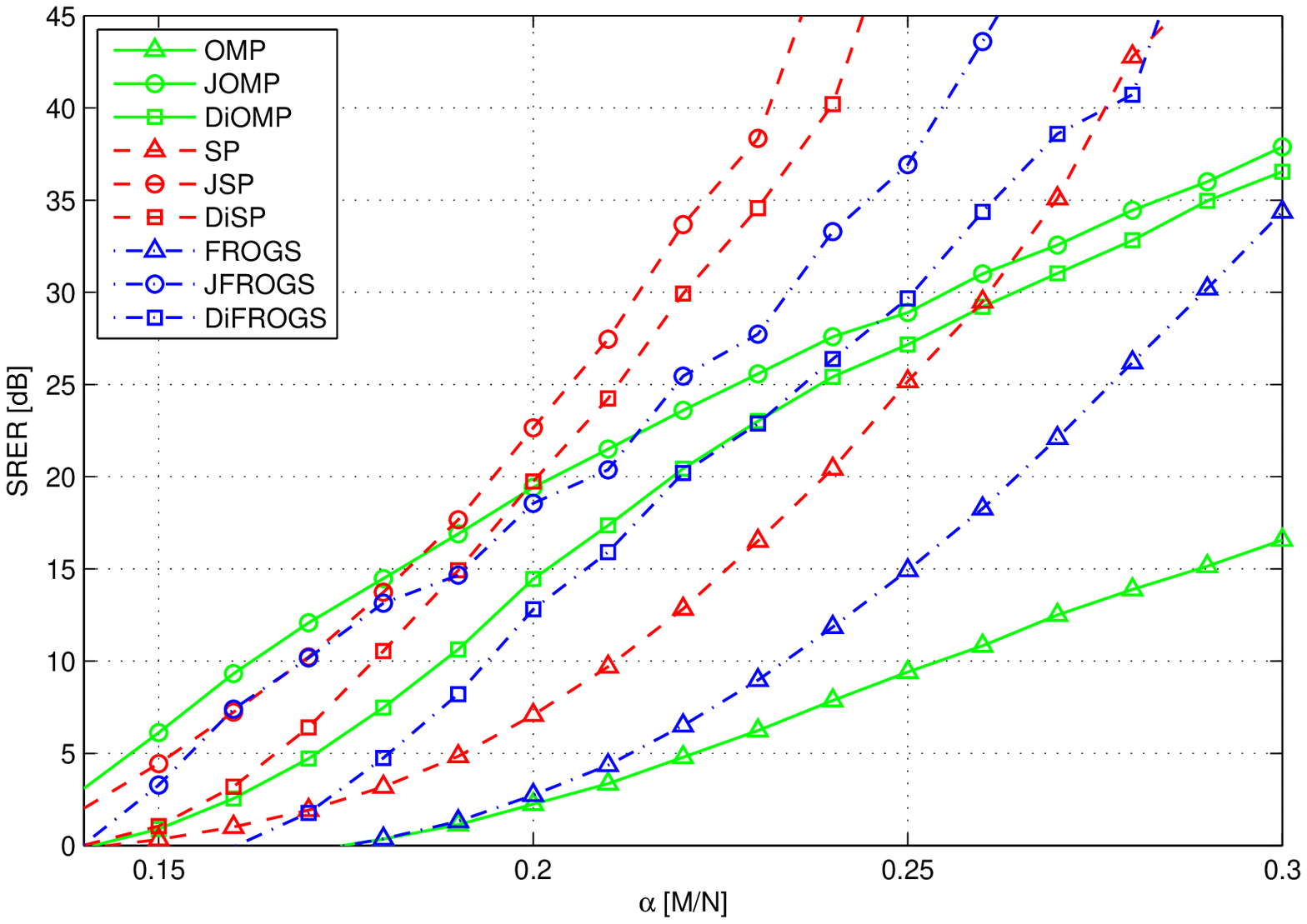}
      \label{fig:srerbin1smnrinf}
    }
  } \qquad
  \subfloat[\asce versus fraction of measurements]{
    \resizebox{0.46\columnwidth}{!}{
      \includegraphics[width=\columnwidth]{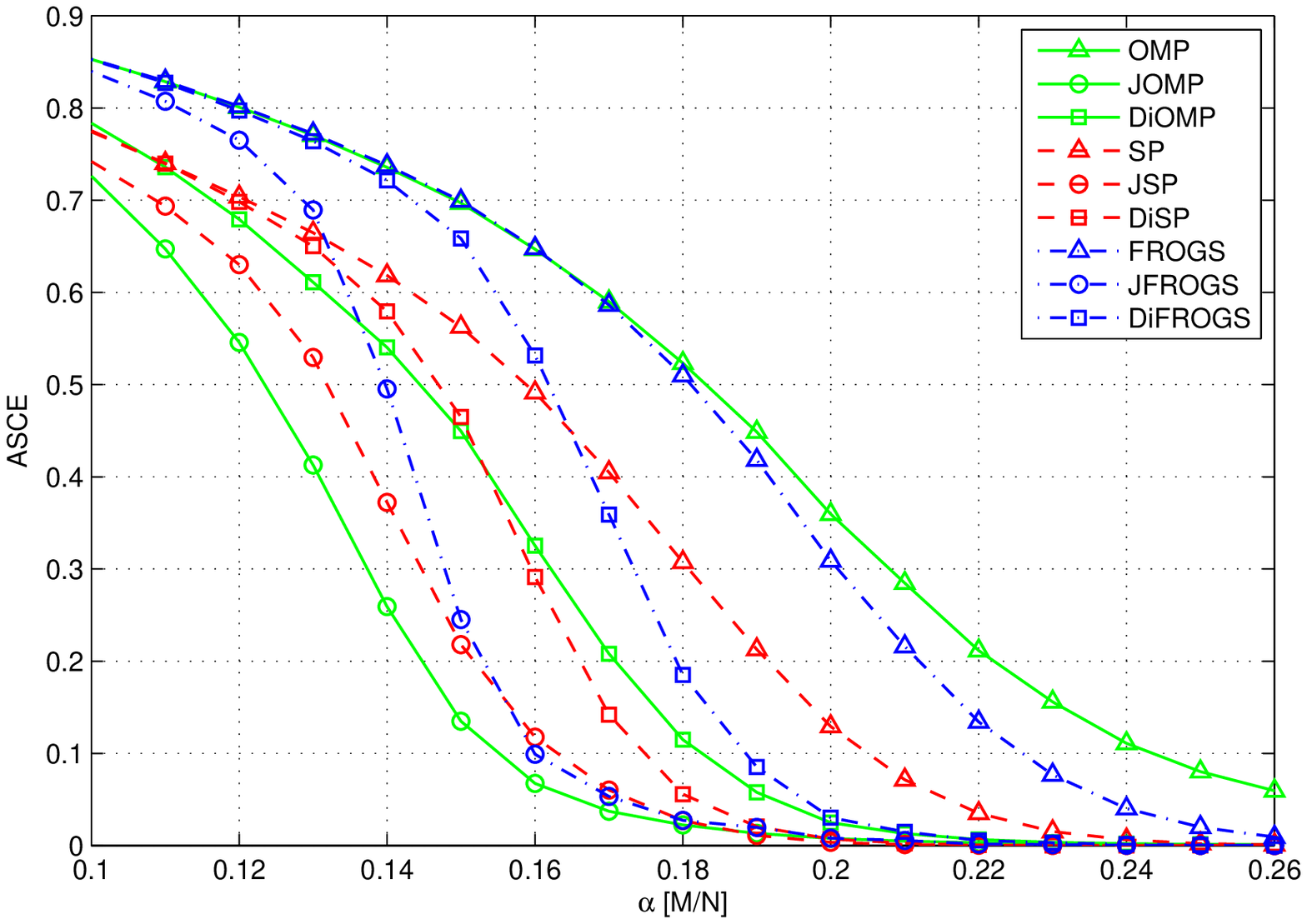}
      \label{fig:ascebin1smnrinf}
    }
  }
  \caption{Comparison of \gp, \digp and joint \gp algorithms for {\textbf{binary}} sparse signal at \textbf{clean measurement} condition. We show performance results against fraction of measurements.}
  \label{fig:bin1smnrinf}
  \subfloat[\srer (in dB) versus fraction of measurements]{
    \resizebox{0.46\columnwidth}{!}{
      \includegraphics[width=\columnwidth]{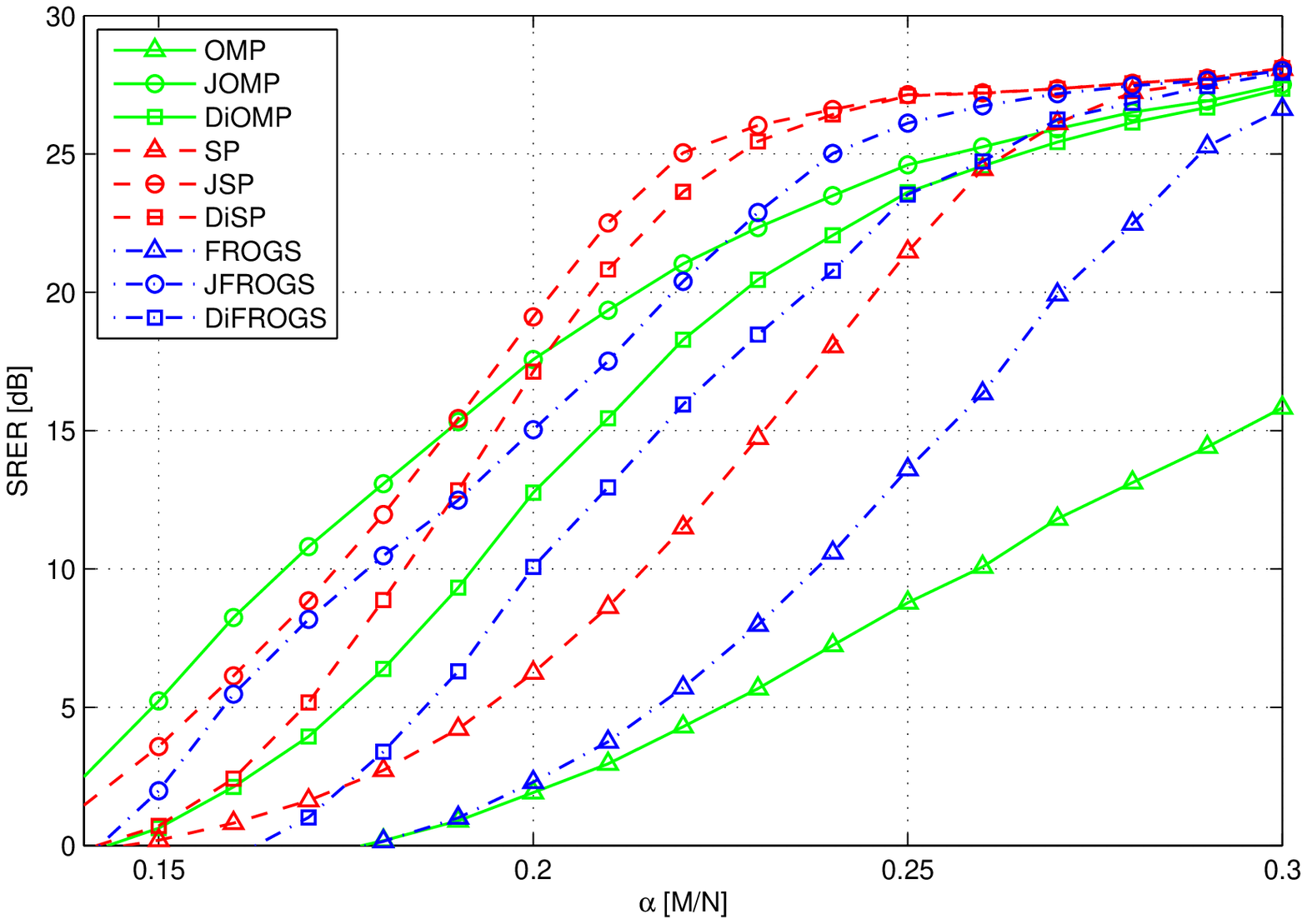}
      \label{fig:srerbin1smnr20}
    }
  } \qquad
  \subfloat[\asce versus fraction of measurements]{
    \resizebox{0.46\columnwidth}{!}{
      \includegraphics[width=\columnwidth]{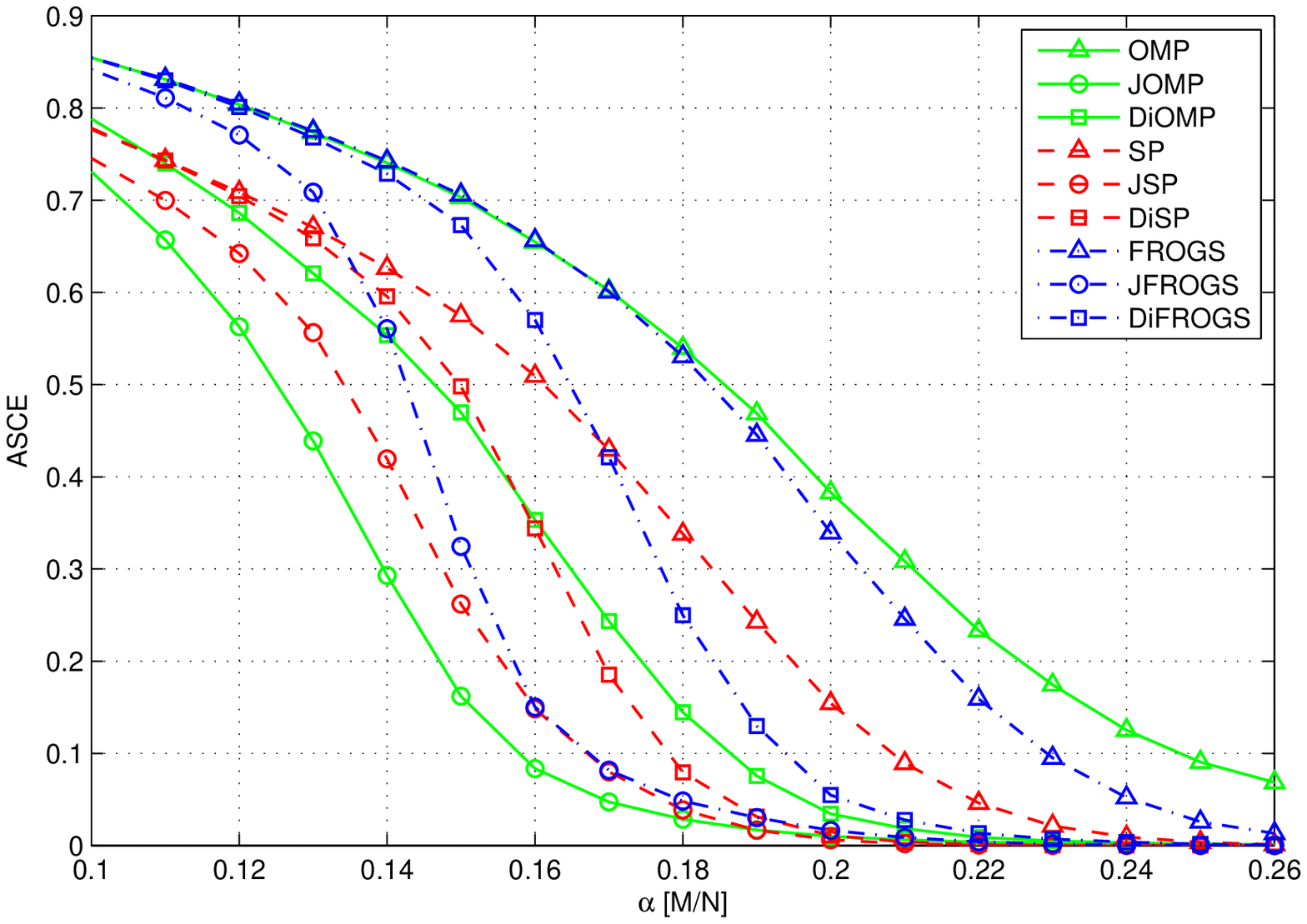}
      \label{fig:ascebin1smnr20}
    }
  }
  \caption{Comparison of \gp, \digp and joint \gp algorithms for {\textbf{binary}} sparse signal at \textbf{noisy measurement} condition, where \smnr = 20 dB. We show performance results against fraction of measurements.}
  \label{fig:bin1smnr20}
\end{figure}
\end{landscape}


%\vfill


\end{document}